\documentclass[amsmath,12pt,amssymb,preprint,prd,aps,nofootinbib]{revtex4}
\usepackage{amsfonts} 
\usepackage{graphicx} 
\usepackage{epsfig}
\usepackage{diagbox}
\usepackage{slashbox}
\usepackage{hyperref}
\usepackage{multirow}
\usepackage{bm}

\begin{document}
\title{Open Strange Mesons in (magnetized) nuclear matter}

\author{Ankit Kumar}
\email{ankitchahal17795@gmail.com}
\author{Amruta Mishra}
\email{amruta@physics.iitd.ac.in}  
\affiliation{Department of Physics, 
Indian Institute of Technology, Delhi, Hauz Khas, New Delhi - 110016}
\begin{abstract}
We investigate the mass modifications of open strange mesons (vector $K^*$ and axial vector $K_1$) in (magnetized) isospin asymmetric nuclear matter using quantum chromodynamics sum rule (QCDSR) approach. The in-medium decay widths of $K^*$ $\rightarrow$ $K\pi$ and $K_1$ $\rightarrow$ $K^*\pi$ are studied from the mass modifications of $K_1$, $K^*$ and $K$ mesons, using a light quark-antiquark pair creation model, namely the ${}^3 P_0$ model. The in-medium decay width for $K_1$ $\rightarrow$ $K^*\pi$ is compared with the decay widths calculated using a phenomenological Lagrangian. The effects of magnetic fields are also studied on the mass and the partial decay width of the vector $K^*$ meson decaying to $K\pi$. Within the QCD sum rule approach, the medium effects on the masses of the open strange mesons are calculated from the light quark condensates and the gluon condensates in the hadronic medium. The quark condensates are calculated from the medium modifications of the scalar fields ($\sigma$, $\zeta$, and $\delta$) in the mean field approximation within a chiral $SU(3)$ model, while the scalar gluon condensate is obtained from the medium modification of a scalar dilaton field ($\chi$), which is introduced within the model to imitate the scale invariance breaking of QCD. 
\end{abstract}
\maketitle
\section{Introduction}
\label{introduction_section}
The investigation of various properties of hadrons \cite{Hosaka_PPNP96_88_2017} has become an important and emerging topic of research interest in high energy physics due of its relevance to relativistic heavy ion collision (HIC) experiments. The medium produced in these heavy ion colliders has high density and/or high temperature, which can affect the experimental observables due to the medium modifications of the produced  hadrons. It is important to study the effects of isospin asymmetry as the colliding heavy ions have more number of neutrons as compared to the number of protons. The study of light vector mesons is important due to its relevance to observables e.g., the dilepton spectra in the HIC experiments. The dileptons are promising observables for the study the properties of hadrons in dense nuclear matter as their interaction with the hadronic environment is negligible, and, they give information about all stages 
of the evolution of the strongly interacting created in heavy ion collision experiments. The in-medium masses of light vector mesons ($\rho$, $\omega$, and $\phi$) have been studied in strange hadronic matter \cite{AM_PRC91_035201_2015} and isospin asymmetric magnetized nuclear medium \cite{AM_PRC100_015207_2019} within the QCD sum rule approach, using the quark and gluon condensates as calculated within a chiral $SU(3)$ model. 
The study of strange mesons has been the center of attention due to their significance in studying the yield and spectra of these mesons, produced in HIC experiments \cite{Hartnack_PR510_119_2012} as well as in the study of certain astronomical bodies where strange matter could exist in the interior of the neutron stars \cite{Tolos_PPNP112_103770_2020,Kaplan_PLB175_57_1986}. The in-medium masses of the open pseudoscalar mesons, e.g., the kaons and antikaons have been studied in strange hadronic matter using a chiral effective model \cite{AM_PRC70_044904_2004} and the effects of temperature have also been incorporated. The effects from baryonic Dirac sea are also investigated in \cite{AM_PRC70_044904_2004} and the results for in-medium masses are compared to that obtained from chiral perturbation theory (ChPT).
The energies and optical potentials for kaons and antikaons have been studied in isospin asymmetric nuclear (hyperonic) matter within a chiral effective model \cite{AM_PRC78_024901_2008, AM_PRC74_064904_2006, AM_EPJA41_205_2009}. The isospin asymmetry effects are observed to be significant at high densities. These can have observable consequences in the heavy-ion beam collisions at Compressed Baryonic Matter (CBM) experiment at future Facility for Antiproton and Ion Research (FAIR). Moreover, the low energy scattering of antikaons with nucleons ($N$) have been studied in the framework of coupled channel approach \cite{Oset_NPA635_99_1998,Ramos_NPA671_481_2000} and the $\Lambda (1405)$ is reproduced dynamically just below the $\Bar{K}N$ threshold due to coupling of $\bar K N$ channel to $\pi \Sigma$ channel in $I=0$ \cite{Koch_PLB337_7_1994}. 

In the present work, we study the in-medium masses as well as the decay widths of vector $K^*$ (decaying to $K\pi$) and axial vector $K_1$ (to $K^*\pi$) mesons, in (magnetized) isospin asymmetric dense nuclear medium. These strange vector and axial-vector mesons are the chiral partners of each other and are considered as obvious systems to study the chiral symmetry breaking effects and its restoration in the medium. In Ref. \cite{Gubler_PLB767_336_2017}, the properties of $f_1(1285)$ and $\omega$ mesons are studied in QCDSR approach to probe the chiral symmetry restoration in nuclear environment.  The in-medium masses for the light vector mesons ($\rho$, $\omega$, and $\phi$) have been calculated within QCD sum rule approach \cite{Hatsuda_PRC46_R34_1992} from the medium modifications of quark condensates and scalar gluon condensates. The lowest Charmonium states $J/\psi$ and $\eta_c$ have been studied in isospin asymmetric hot nuclear matter \cite{AK_PRC82_045207_2010} within QCD sum rule approach, and the effects of medium density are found to be the dominant effects. Moreover, within the QCD sum rule approach, the masses of $1S$ and $1P$ states of heavy quarkonium (charmonium and bottomonium) have been studied in isospin asymmetric nuclear matter including the effects of strong magnetic fields \cite{Parui_PRD106_114033_2022}. The mass modification for heavy quarkonium states, within QCD sum rule approach, arises from the medium modifications of scalar gluon condensates and twist-2 gluon condensates. This work includes the effects from Dirac sea through summing over the nucleonic tadpole diagrams and leads to a decrease in the values of light quark condensates with magnetic field, an effect known as inverse magnetic catalysis, when the anomalous magnetic moments (AMMs) of nucleons are taken into account. The decay width of channel  $K^*$ $\rightarrow$ $K\pi$ is studied using the ${}^3 P_0$ model, from the in-medium masses of vector $K^*$ and pseudoscalar $K$ meson, which are calculated within QCD sum rule approach and chiral $SU(3)$ model, respectively. The open flavor mesons decay through the creation of a $\bar{q}q$ pair which is produced with vacuum quantum numbers ($J^{PC}=0^{++}$) corresponding to a  ${}^3 P_0$ state \cite{Ackleh_PRD54_6811_1996,Micu_NPB10_521_1969}. The ${}^3 P_0$ model has been used extensively in the literature to study the decays of various mesons \cite{Yaouanc_PRD8_2223_1973,Barnes_PRD55_4157_1997,Friman_PLB548_153_2002}. This model indicates the importance of taking into account the internal structures of hadrons as it has explained the experimentally observed suppression of $\psi(4040)$ decay mode to $\bar{D}D$ and ($\bar{D^*}D$ + $\bar{D}D^*$) in comparison with $\bar{D^*}D^*$ decay mode \cite{Yaouanc_PLB71_397_1977}.

The masses of strange mesons ($K$, $K^*$, and $\phi$) have been investigated in the presence of strong magnetic fields in Ref. \cite{AM_IJMPE30_2150014_2021} due to $\phi-\eta^\prime$ and $K^*-K$ mixing, and also the decay widths for $\phi \rightarrow K \bar{K}$, $K^* \rightarrow K\pi$ are studied in a field theoretic model for composite hadrons from the mass modifications of the mesons. The in-medium spectral functions and production cross-sections for the strange mesons ($K^*, \Bar{K}^*$, and $\phi$), in strange hadronic medium, have also been studied from the in-medium masses and decay widths for these mesons \cite{AM_EPJA57_98_2021}. The effects of medium density as well as strangeness on the production cross-sections of $K^*, \Bar{K}^*$, and $\phi$ from the $K \pi$, $\Bar{K} \pi$, and $\Bar{K}K$ scattering respectively, have been found to be quite appreciable when compared to the vacuum conditions. The properties of vector $\bar{K^*}$ meson in the nuclear matter are investigated in Ref. \cite{Tolos_PRC82_045210_2010} using a unitary approach in coupled channels. The strange vector $K^*$($\bar{K^*}$) mesons are produced mainly at later stages from the $K\pi(\bar{K}\pi$) scattering and the contribution from direct hadronization from the quark gluon plasma (QGP) state is quite small, as calculated within Parton–Hadron-String-Dynamics (PHSD) transport model \cite{Ilner_PRC95_014903_2017, Ilner_PRC99_024914_2019}. Although the study of medium density effects might find relevance in future experiments at the GSI Facility for Antiproton and Ion Research (FAIR) and Nuclotron-based Ion Collider facility (NICA) \cite{Kumar_EPJC79_403_2019,Rapp_PNP65_209_2010}, where matter having large baryon density will be produced. The axial vector $K_1$ meson masses have been analyzed in the QCD sum rule analysis \cite{Song_PLB792_160_2019} in the nuclear matter and the decay widths for $K_1 \rightarrow K\pi\pi$ channel have been studied using the ${}^3 P_0$ model \cite{Tayduganov_PRD85_074011_2012}. 

 In QCD sum rule approach \cite{AM_PRC91_035201_2015,Hatsuda_PRC46_R34_1992,Hatsuda_PRC52_3364_1995}, we expand the current-current correlation function for the corresponding meson using operator product expansion (OPE) in terms of local operators and their coefficients. The central idea of this approach is to relate the spectral density of this correlation function with the OPE expression via a dispersion relation, for large space-like regions. The medium modifications in masses arise due to the medium modifications of light quark condensates and gluon condensates within the QCD sum rule approach. These light quark condensates are related to scalar fields ($\sigma$, $\zeta$, and $\delta$) of the medium by comparing the explicit chiral symmetry breaking term of QCD Lagrangian to the corresponding Lagrangian term in the chiral $SU(3)$ model \cite{Papazoglou_PRC59_411_1999,Zschiesche_PRC63_025211_2001}; while the gluon condensates are related to the scalar dilaton field ($\chi$) of the medium. The chiral effective Lagrangian is written such that it includes various symmetries of low energy QCD and the symmetry breaking effects. The coupled equations of motion for various scalar fields ($\sigma$, $\zeta$, $\delta$, and $\chi$) are solved within the chiral $SU(3)$ model including various medium effects. 
 
The estimation of strong magnetic field production in the peripheral HIC experiments \cite{Tuchin_AHEP2013_490495_2013,Skokov_IJMPA24_5925_2009} have grown immense interest in the study of the magnetic field effects on the produced medium. The estimated magnetic field strength at Relativistic Heavy Ion Collider (RHIC) is $\sim$ $2 m_\pi^2$ and at Large Hadron Collider (LHC) is $\sim$ $15 m_\pi^2$, calculated considering Lienard-Wiechert potential in numerical simulations within a microscopic transport model \cite{Skokov_IJMPA24_5925_2009}. The study of strong magnetic field effects on produced medium is also important due to novel interesting quantum effects like chiral magnetic effect \cite{Fukushima_PRD78_074033_2008}, magnetic catalysis \cite{Kharzeev_Springer_2013, Shovkovy_LNP871_13_2013} and inverse magnetic catalysis \cite{Kharzeev_Springer_2013,Preis_LNP871_49_2013} as well as in the study of neutron stars and magnetars where large magnetic fields are estimated to exist. The charged mesons have contribution due to Landau level quantization in the presence of magnetic field and the effects of PV mixing, considered through effective Lagrangian vertex \cite{Gubler_PRD93_054026,Cho_PRD91_045025_2015,AM_PRC102_045204_2020,AM_IJMPE30_2150064_2021} and spin-magnetic field interaction term \cite{Alford_PRD88_105017_2013,Yoshida_PRD94_074043_2016,Machado_PRD88_034009_2013}, are also investigated in this study. In Ref. \cite{AM_PRC102_045204_2020}, the PV mixing between vector and pseudoscalar charmonium states is considered through the effective Lagrangian vertex in the presence of strong magnetic fields and the decay width of vector charmonium state $\psi(3770)$ to $\Bar{D}D$ is also studied within a field theoretic model of composite hadrons. Furthermore, the PV mixing between open charm mesons ($D^*$ and $D$) is also studied in Ref. \cite{AM_IJMPE30_2150064_2021} alongwith the  contribution due to Landau levels for charged mesons, and the decay widths ($D^* \rightarrow D \pi$) and ($\psi(3770) \rightarrow \Bar{D} D$) are also studied using a field theoretic model. In field theoretic model of composite hadrons, the hadronic states like charmonium ($\psi(3770)$) state, open charm mesons ($D^*, \bar{D}, D$) and pion ($\pi$) states are constructed explicitly from constituent quark fields assuming harmonic oscillator wave functions for these states, and the matrix element for the decay is then calculated from the light quark-antiquark pair creation term of the free Dirac Hamiltonian density. However, the produced medium density will be very small in these peripheral collision experiments. Therefore, to understand the behavior at these conditions of high magnetic field and low medium density, we also study the effects of high magnetic fields on the properties of vector $K^*$ meson. Hence, this present work might have relevance in lower energy central collisions, where produced medium has high density, as well as in high energy peripheral collision experiments where large magnetic fields are produced but the produced medium has low density.  

The present work is organized in the following manner: In section \ref{SU3_model_section}, we briefly describe the chiral $SU(3)$ model to compute the medium modifications of the quark condensates and scalar gluon condensates from the medium modifications of the scalar fields ($\sigma$, $\zeta$, $\delta$, and $\chi$). In section \ref{QCDSR_section}, we discuss the QCD sum rule approach, which is used to study the in-medium masses of these open strange mesons in isospin asymmetric nuclear medium. We also discuss the effects of strong magnetic field on the masses (and hence the decay widths) of the open strange mesons. In section \ref{self_energy_section}, we discuss the in-medium masses and decay width of $K^*$ meson within a phenomenological model. The decay width of $K_1 \rightarrow K^* \pi$ is also studied within a phenomenological Lagrangian approach. In section \ref{DW_section}, we briefly discuss the ${}^3 P_0$ model, which will be further used to calculate the in-medium decay width of vector $K^*$ and axial vector $K_1$ mesons. In section \ref{result_section}, we discuss and analyze the results obtained and compare them with earlier work to emphasize on the relevance of this work. In section \ref{summary_section}, we summarize the results obtained in the present work.

\section{The Hadronic Chiral $SU(3)$ Model}
\label{SU3_model_section}
We make use of an effective chiral $SU(3)$ model \cite{Papazoglou_PRC59_411_1999,Zschiesche_PRC63_025211_2001,AM_PRC69_024903_2004,Weinberg_PR166_1568_1968,Coleman_PR177_2239_1969,Bardeen_PR177_2389_1969} to calculate the quark condensates and scalar gluon condensates in the nuclear matter, which will further be used in the QCD sum rule approach to calculate the in-medium masses of vector $K^*$ and axial vector $K_1$ mesons. The effective chiral model is formulated on the basis of nonlinear realization of chiral symmetry of QCD and its scale invariance breaking \cite{AM_PRC69_024903_2004,Weinberg_PR166_1568_1968,Coleman_PR177_2239_1969,Bardeen_PR177_2389_1969}. The broken scale invariance of QCD symmetry leads to trace anomaly of QCD, $\theta_\mu^{\mu}=\big<\frac{\beta_{QCD}}{2g}G_{\mu\nu}^{a}G^{a\mu\nu}\big>$, where $G^a_{\mu\nu}$ is the gluon field strength tensor. At the tree level, this is introduced in the effective Lagrangian density through a logarithmic scale breaking term given by \cite{Schechter_PRD21_3393_1980}, 
\begin{eqnarray}
\mathcal{L}_{\rm scale\,breaking}=-\frac{1}{4} \chi^4 {\rm ln}\left(\frac{\chi^4}{\chi_0^4}\right)+\frac{d}{3} \chi^4 {\rm ln}\left(\Big(\frac{\sigma^2 \zeta}{\sigma_0^2\zeta_0}\Big)\Big(\frac{\chi}{\chi_0}\Big)^3\right)
\label{scale_break}
\end{eqnarray}
  
and the total Lagrangian density for the dilaton field ($\chi$) is given by,
 \begin{eqnarray}
\mathcal{L}_{\chi}=\frac{1}{2}\big(\partial_\mu\chi\big)\big(\partial^\mu\chi\big)-k_4\chi^4 + \mathcal{L}_{\rm scale\,breaking}
\label{L_chi}
 \end{eqnarray}
   
where first term is the kinematic term and the second term is introduced to ensure the vacuum expectation value (VEV) for scalar dilaton field.  Then the energy momentum tensor in chiral model is given by,
 \begin{eqnarray}
T_{\mu\nu}=\big(\partial_\mu\chi\big)\left(\frac{\partial\mathcal{L}_{\chi}}{\partial\big(\partial^\nu\chi\big)}\right)-g_{\mu\nu}\mathcal{L}_\chi
\label{T_mu_nu}
 \end{eqnarray}
 
 On the other hand, the energy momentum tensor in QCD \cite{Lee_P72_97_2009, Morita_PRC77_064904_2008}, accounting for the current quark masses, is written as, 
 \begin{eqnarray}
 T_{\mu\nu}= -ST\big(G^a_{\mu\sigma}G^{a\sigma}_{\nu}\big)+\frac{g_{\mu\nu}}{4}\Big(\sum_{i}m_i\big<\bar{q_i}q_i\big>+\Big<\frac{\beta_{QCD}}{2g}G_{\sigma\kappa}^{a}G^{a\sigma\kappa}\Big>\Big)
 \label{T_mu_nu_QCD}
 \end{eqnarray}
 
 where the first term represents the symmetric traceless part and the second term contains the trace part. After multiplying eq. (\ref{T_mu_nu}) and (\ref{T_mu_nu_QCD}) by $g^{\mu \nu}$, we get the trace ($T_\mu^\mu$) of the energy momentum tensor in chiral model and QCD respectively, and the expression for scalar gluon condensate is then given as, 
\begin{eqnarray} 
\Big<\frac{\alpha_s}{\pi}G^a_{\mu\nu}G^{a\mu\nu}\Big>=\frac{8}{9}\Big[\big(1-d\big)\chi^4+\sum_{i} m_i \big<\bar{q_i}q_i\big>\Big]
\label{G_mu_nu}
\end{eqnarray}
where the one-loop QCD $\beta$-function is given by,
 \begin{equation}
\beta_{QCD}(g)= -\frac{11 N_c g^3}{48\pi^2}\Big(1-\frac{2N_f}{11N_c}\Big)+\mathcal{O}\big(g^5\big)
\label{beta_QCD}
\end{equation}
with $N_c=3$, and $N_f=3$, are the number of colors and quark flavors respectively and the strong coupling constant of QCD, $\alpha_s = g^2/4\pi$. Thus, the scalar gluon condensate is introduced in the chiral $SU(3)$ model through a scalar dilaton field ($\chi$). The scalar gluon condensate has additional contribution due to finite non-zero quark masses $m_i(i = u,d,s)$. The non-zero quark condensates $\big<\Bar{q}_i q_i\big>$ are introduced in the QCD vacuum by the spontaneous breaking of chiral symmetry by the ground state. The finite quark mass term \Big($m_i\big<\bar{q_i}q_i\big>$\Big) of equation (\ref{G_mu_nu}) is given in terms of scalar fields \big($\sigma$, $\zeta,$ and $\delta$\big) by comparing the explicit chiral symmetry breaking term of the chiral model, after mean field approximation,
\begin{equation}
\mathcal{L}_{SB}={\rm Tr}\left[{\rm diag}\,\Big(-\frac{1}{2}m_\pi^2 f_\pi\big(\sigma+\delta\big),-\frac{1}{2}m_\pi^2 f_\pi\big(\sigma-\delta\big),\Big(\sqrt{2}m_K^2 f_K-\frac{1}{\sqrt{2}}m_\pi^2 f_\pi\Big)\zeta\Big)\right]
\label{L_SB}
\end{equation}
with the corresponding Lagrangian density term of the QCD, which is written as,
\begin{equation}
\mathcal{L}_{SB}^{QCD}=-{\rm Tr}\left[{\rm diag}\,\big(m_u \bar{u}u,m_d \bar{d}d,m_s \bar{s}s\big)\right]
\label{L_SB_QCD}
\end{equation}

Within the chiral model, the coupled equations of motion for the scalar fields \big($\sigma$, $\zeta$, $\delta$ and $\chi$\big), derived from the chiral Lagrangian density, are given as,
\begin{eqnarray}
&k_0\chi^2\sigma - 4k_1 \big(\sigma^2+\zeta^2+\delta^2\big)\sigma-2k_2\big(\sigma^3+3\sigma\delta^2\big)-2k_3\chi\sigma\zeta\nonumber\\& -\frac{d}{3}\chi^4\left(\frac{2\sigma}{\sigma^2-\delta^2}\right)+\left(\frac{\chi}{\chi_0}\right)^2 m_\pi^2 f_\pi-\sum_{i} g_{\sigma i}\rho_i^s=0
\label{sigma_field}
\end{eqnarray}
\begin{eqnarray}
&k_0\chi^2\zeta-4k_1\big(\sigma^2+\zeta^2+\delta^2\big)\zeta-4k_2\zeta^3-k_3\chi\big(\sigma^2-\delta^2\big)-\frac{d}{3}\frac{\chi^4}{\zeta}\nonumber 
\\&+\left(\frac{\chi}{\chi_0}\right)^2\left[ \sqrt{2}m_K^2 f_K - \frac{1}{\sqrt{2}} m_\pi^2 f_\pi\right]-\sum_{i} g_{\zeta i}\rho_i^s=0
\label{zeta_field}
\end{eqnarray}
\begin{eqnarray}
&k_0\chi^2\delta-4k_1\big(\sigma^2+\zeta^2+\delta^2\big)\delta-2k_2\big(\delta^3+3\sigma^2\delta\big)+ 2k_3\chi\delta\zeta\nonumber\\&+\frac{2}{3}d\chi^4\left(\frac{\delta}{\sigma^2-\delta^2}\right)-\sum_{i} g_{\delta i}\rho_i^s=0
\label{delta_field}
\end{eqnarray}
\begin{eqnarray}
&k_0\chi\big(\sigma^2+\zeta^2+\delta^2\big)-k_3\big(\sigma^2-\delta^2\big)\zeta+\chi^3\left[1+{\rm ln}\Big(\frac{\chi^4}{\chi_0^4}\Big)\right]+ \nonumber\\& \big(4k_4-d\big)\chi^3-\frac{4}{3}d \chi^3{\rm ln}\left(\Big(\frac{(\sigma^2-\delta^2)\zeta}{\sigma_0^2\zeta_0}\Big)\Big(\frac{\chi}{\chi_0}\Big)^3\right) + \bigg(\frac{2\chi}{\chi_0^2}\bigg)\nonumber\\ & \left[\Big(m_\pi^2 f_\pi\sigma+\big(\sqrt{2}m_K^2 f_K-\frac{1}{\sqrt{2}}m_\pi^2 f_\pi\big)\zeta\Big)\right]=0
\label{chi_field}
\end{eqnarray}

The medium effects due to the baryon density, isospin asymmetry, and magnetic field, are incorporated into the model through the scalar fields, which depend on the scalar densities ($\rho_i^s$) of the baryons. The coupled equations of motion, given by (\ref{sigma_field},\ref{zeta_field},\ref{delta_field},\ref{chi_field}), are solved to find the medium dependent values of the scalar fields,
from which we obtain the scalar gluon condensates $\big<\frac{\alpha_s}{\pi}G^a_{\mu\nu}G^{a\mu\nu}\big>$ and the quark condensates \Big($\big<\bar{u}u\big>$, $\big<\bar{d}d\big>$, and $\big<\bar{s}s\big>$\Big) in the nuclear medium.

\section{QCD Sum Rule Approach}
\label{QCDSR_section}
In this section, we will discuss the QCD sum rule method \cite{AM_PRC91_035201_2015,Song_PLB792_160_2019,Klingl_NPA624_527_197}, which is used to calculate the in-medium masses through the medium modifications of the quark and gluon condensates, calculated within the chiral model. The current-current correlation function, written in terms of the time-ordered product of two currents, for the meson $V$ is given by, 
\begin{equation}
\Pi^{V}_{\mu\nu}(q)=i\int d^4x\,e^{iq.x}\big<0\big|T\big[j^{V}_{\mu}(x),j^{V}_{\nu}(0)\big]\big|0\big>
\label{Correlator_1}
\end{equation}
where the currents for vector $K^*$ meson are given as $j^{K^{*+}}_{\mu}= \bar{s}\gamma_{\mu}u$ and $j^{K^{*0}}_{\mu}= \bar{s}\gamma_{\mu}d$; while the currents for the axial vector $K_1$ meson are given by $j^{K_1^+}_{\mu}= \bar{s}\gamma_{\mu}\gamma_{5}u$ and $j^{K_1^0}_{\mu}= \bar{s}\gamma_{\mu}\gamma_{5}d$. We write the transverse tensor structure for the correlation function as a sum of two independent functions as \cite{Song_PLB792_160_2019,Leupold_PRC64_015202_2001}, 
\begin{equation}
\Pi^{V}_{\mu\nu}(q)=-g_{\mu\nu}\Pi_{1}(q^2)+q_\mu q_\nu \Pi_2(q^2)
\label{Correlator_2}
\end{equation}
 For the conserved vector current $j^{K^*}_{\mu}$, these two functions are related as $\Pi_{1}(q^2)= q^2 \Pi_2(q^2)$. As the axial current $j^{K_1}_{\mu}$ is not conserved, this relation does not hold true for axial current. We can make use of either $\Pi_{1}(q^2)$  or $\Pi_2(q^2)$ to carry out QCDSR, but $\Pi_2(q^2)$ will have contributions from the pseudoscalar mesons which require further investigation of the in-medium properties of pseudoscalar mesons. Therefore we will make use of $\Pi_1(q^2)$ in this work throughout. The main idea of QCDSR is to relate the spectral density of the correlator function $\Pi^{V}(q^2=-Q^2)$ on the phenomenological side via a dispersion relation with the QCD operator product expansion (OPE) side. The correlator function, on the phenomenological side, can be written as,
\begin{equation}
12\pi^2\widetilde{\Pi}^{V}_{\rm phen}(Q^2)=\int ds \frac{R^{V}_{\rm phen}(s)}{s+Q^2} 
\label{Correlator_phen}
\end{equation}

where $\widetilde{\Pi}^{V}(Q^2)=\Pi^{V}(Q^2)/Q^2$ and the spectral density $R^{V}_{\rm phen}(s)$ is related to the imaginary part of the correlator as $R^{V}_{\rm phen}(s)=12\pi Im(\Pi^{V}_{\rm phen}(s))$. For enhancing the contribution of the pole, we make use of Borel transformation \cite{Shifman_NPB147_385_1979,Reinders_PR127_1_1985} and we get, 
\begin{equation}
12 \pi^2\,\widetilde{\Pi}^{V}(M^2)=\int ds e^{-\frac{s}{M^2}}R^{V}_{\rm phen}(s) 
\label{Correlator_phen_Borel}
\end{equation}
For large space-like regions, $Q^2=-q^2>>1$ GeV$^2$, the asymptotic freedom in QCD allows for series expansion of correlation function in terms of operator product expansion (OPE) as \cite{Klingl_NPA624_527_197,Kwon_PRC81_065203_2010}, 
\begin{equation}
12\pi^2\,\widetilde{\Pi}^{V}(q^2=-Q^2)=d_{V} \bigg[-c_0^{{V}}{\rm ln}\frac{Q^2}{\mu^2}+\frac{c_1^{V}}{Q ^2}+\frac{c_2^{V}}{Q^4}+\frac{c_3^{V}}{Q^6}+...\bigg] 
\label{Correlator_OPE}
\end{equation}
where $d_{K^*,K_1}=3$ and the scale $\mu$ is taken to be 1 GeV here. The first term is the leading perturbative QCD term and subsequent higher order terms, containing the non-perturbative effects of QCD, are suppressed by powers of $1/Q^2$. The coefficients $c_i's$ ($i=2,3...$) are related to quark condensates and scalar gluon condensates. For the $K^{*+}$ meson \cite{Song_PLB792_160_2019,Shifman_NPB147_385_1979,Shifman_NPB147_448_1979}, these coefficients are given by,
\begin{equation}
c_0^{K^*}=1+\frac{\alpha_s(Q^2)}{\pi},\,\, c_1^{K^*}=-3(m_u^2+m_s^2)
\label{c0_c1_K*}
\end{equation}
\begin{equation}
c_2^{K^*}=\frac{\pi^2}{3}\Big<\frac{\alpha_s}{\pi}G^{\mu\nu}G_{\mu\nu}\Big>+\frac{16\pi^2}{d_{K^*}}\big<m_u\bar{s}s+m_s\bar{u}u\big>-\frac{4\pi^2}{d_{K^*}}\big<m_u\bar{u}u+m_s\bar{s}s\big>
\label{c2_K*}
\end{equation}
\begin{align}
& c_3^{K^*} =-8\pi^3\Bigg[2\big<\alpha_s(\bar{u}\gamma_{\mu}\gamma_{5}\lambda^a s)(\bar{s}\gamma^{\mu}\gamma^{5}\lambda^a u)\big> + \frac{2}{9}\Big<(\bar{u}\gamma_{\mu}\lambda^a u+ \bar{s}\gamma_{\mu}\lambda^a s)\times\nonumber\\&\Big(\sum_{q=u,d,s}\bar{q}\gamma^{\mu}\lambda^a q\Big)\Big>\Bigg]=-8\pi^3\alpha_s\kappa_{n}\bigg[\frac{32}{9}\big<\bar{u}u\big>\big<\bar{s}s\big> + \frac{32}{81}\big(\big<\bar{u}u\big>^2+ \big<\bar{s}s\big>^2\big)\bigg] 
\label{c3_K*}
\end{align}

where $\alpha_s(Q^2)=4\pi/\big[b\,      {\rm ln}(Q^2/\Lambda^2_{QCD})\big]$ is the running coupling constant of QCD and $(m_u,m_s)$ are current quark masses for up and strange quark. The QCD scale is taken to be $\Lambda_{QCD}=140$ MeV, with $b=11-(2/3)N_f$ with $N_f=3$ as the number of quark flavors. To evaluate the four quark operators of coefficient $c_3^{K^*}$, we have used the factorization method \cite{Shifman_NPB147_448_1979},
\begin{equation}
\big<(\bar{q_i}\gamma_{\mu}\gamma_{5}\lambda^a q_j)^2\big> = - \big<(\bar{q_i}\gamma_{\mu}\lambda^a q_j)^2\big>= \delta_{ij} \frac{16}{9}\kappa_n\big<\bar{q_i}q_i\big>^2
\label{factorization_method}
\end{equation}

where $q_i = (u,d, s)$ and the parameter $\kappa_n$ is introduced to parameterize the deviation from the exact factorization with $`n$' corresponding to different mesons. For the neutral $K^{*0}$ meson, the up ($u$) quark flavor is replaced by the down ($d$) quark flavor. The coefficients for the strange axial vector meson $K_1^+$ are given by \cite{Song_PLB792_160_2019},

\begin{equation}
c_0^{K_1}=1+\frac{\alpha_s(Q^2)}{\pi},\,\, c_1^{K_1}=-3(m_u^2+m_s^2)
\label{c0_c1_K1}
\end{equation}
\begin{equation}
c_2^{K_1}=\frac{\pi^2}{3}\Big<\frac{\alpha_s}{\pi}G^{\mu\nu}G_{\mu\nu}\Big>-\frac{16\pi^2}{d_{K_1}}\big<m_u\bar{s}s+m_s\bar{u}u\big>+\frac{4\pi^2}{d_{K_1}}\big<m_u\bar{u}u+m_s\bar{s}s\big> 
\label{c2_K1}
\end{equation}
\begin{align}
& c_3^{K_1} =-8\pi^3\Bigg[2\big<\alpha_s(\bar{u}\gamma_{\mu}\lambda^a s)(\bar{s}\gamma^{\mu}\lambda^a u)\big> + \frac{2}{9}\Big<(\bar{u}\gamma_{\mu}\lambda^a u+ \bar{s}\gamma_{\mu}\lambda^a s)\times\nonumber\\&\Big(\sum_{q=u,d,s}\bar{q}\gamma^{\mu}\lambda^a q\Big)\Big>\Bigg]=-8\pi^3\alpha_s\kappa_n\bigg[-\frac{32}{9}\big<\bar{u}u\big>\big<\bar{s}s\big> + \frac{32}{81}\big(\big<\bar{u}u\big>^2+ \big<\bar{s}s\big>^2\big)\bigg] 
\label{c3_K1}
\end{align}

Thus the difference in correlator function for the vector and axial vector is proportional to chiral symmetry breaking dimension-4 \big(\big<$m_{q_i}\bar{q_i}q_i$\big>\big) and dimension-6 \big($\big<\bar{q_i}q_i\big> \big<\bar{q_j}q_j\big>$\big) operators. After doing the Borel transformation of equation (\ref{Correlator_OPE}) for improved convergence, we get,
\begin{equation}
12 \pi^2\,{\widetilde{\Pi}^{V}(M^2})=d_{V}\bigg[c_0^{V} M^2+c_1^{V}+\frac{c_2^{V}}{M^2}+\frac{c_3^{V}}{2M^4}\bigg] 
\label{Correlator_OPE_Borel}
\end{equation}
We assume that the spectral density has a resonance part $R^{V({\rm res})}_{\rm phen}(s)$ and a perturbative continuum that contains all higher energy states, which are separated by an energy scale $\sqrt{s_0^{V}}$ as \cite{Klingl_NPA624_527_197,Kwon_PRC81_065203_2010},
\begin{equation}
R^{V}_{\rm phen}(s)=R^{{V}({\rm res})}_{\rm phen}(s)\, \Theta(s_0^{V}-s)+d_{V}c_0^{V}\,\Theta(s-s_0^{V})
\label{R_phen}
\end{equation}

As the rapid crossover between the resonance and continuum part is not realistic, the scale $s_0^V$ is taken as the average scale characterizing the smooth transition from low-lying resonance region to high-energy continuum part. Due to larger exponential suppression of the correlator function through utilizing the Borel transform, the more detailed description of the crossover and continuum region becomes insignificant \cite{Leupold_PRC64_015202_2001}. The correlator function is matched from equations (\ref{Correlator_phen_Borel}) and (\ref{Correlator_OPE_Borel}) to get,
\begin{equation}
\int ds\,e^{-\frac{s}{M^2}}R^{V}_{\rm phen}(s)= d_{V}\bigg[c_0^{V} M^2+c_1^{V}+\frac{c_2^{V}}{M^2}+\frac{c_3^{V}}{2M^4}\bigg]
\label{matching_phen_OPE}
\end{equation}

 We expand the exponential expression, for $M>\sqrt{s_0^{V}},$ of equation (\ref{matching_phen_OPE}) in powers of $\frac{s}{M^2}$ for $s<s_0^V$ and get the finite energy sum rules (FESRs) \cite{AM_PRC91_035201_2015,Klingl_NPA624_527_197} in vacuum, by comparing the coefficients of various powers of $1/M^2$. Using simple ansatz for the vector $K^*$ meson spectral function $R^{K^*}_{\rm phen}(s)$ as \cite{AM_PRC91_035201_2015,Klingl_NPA624_527_197,Kwon_PRC81_065203_2010},
\begin{equation}
R^{K^*}_{\rm phen}(s)=F_{K^*}\,\delta(s-m_{K^*}^2)+d_{K^*}c_0^{K^*}\,\Theta(s-s_0^{K^*})
 \label{simple_ansatz_K*}
\end{equation}
the finite energy sum rules (FESRs) for the vector $K^*$ meson in vacuum are given as,
 \begin{equation}
 F_{K^*}=d_{K^*}(c_0^{K^*} s_0^{{K^*}}+c_1^{K^*}) 
  \label{FESR_K*_vac1}
\end{equation}
\begin{equation}
F_{K^*} \,m_{K^*}^2 = d_{K^*} \Big(\frac{(s_0^{{K^*}})^2 c_0^{K^*}}{2}-c_2^{{K^*}}\Big)
\label{FESR_K*_vac2}
\end{equation}
\begin{equation}
 F_{K^*} \,m_{K^*}^4 = d_{K^*}\Big(\frac{(s_0^{K^*})^3 c_0^{K^*}}{3}+c_3^{K^*}\Big) 
 \label{FESR_K*_vac3}
\end{equation}
These equations are solved by putting in the vacuum masses and vacuum values of condensates to find the delineation scale $s_0^{K^*}$, the overlap strength $F_{K^*}$ between the current and lowest-lying resonance, and the coefficient $c_3^{K^*}$ for vector meson $K^{*+}$ and $K^{*0}$ separately. The parameter $\kappa_n$ can be fixed from the value of coefficient $c_3^{K^*}$. 

However, in the presence of the nuclear medium, the quark and gluon condensates are also modified due to medium modifications of quark and gluon condensates, and these medium effects are incorporated in the FESRs through the medium-modified coefficients $c_2^*$ and $c_3^*$. Then the finite energy sum rules for the $K^*$ meson in the nuclear medium are given by,
\begin{equation}
 F_{K^*}^*=d_{K^*}(c_0^{K^*} s_0^{*{K^*}}+c_1^{K^*})
 \label{FESR_K*_med1}
\end{equation}
\begin{equation}
F_{K^*}^* \,m_{K^*}^{*2} = d_{K^*}\Big(\frac{(s_0^{*{K^*}})^2 c_0^{K^*}}{2}-c_2^{*{K^*}}\Big)
\label{FESR_K*_med2}
\end{equation}
\begin{equation}
 F_{K^*}^* \,m_{K^*}^{*4} = d_{K^*}\Big(\frac{(s_0^{*{K^*}})^3 c_0^{K^*}}{3}+c_3^{*{K^*}}\Big) 
 \label{FESR_K*_med3}
\end{equation}
These FESRs are solved to find the in-medium scale $s_0^{*K^*}$, overlap strength $F^*_{K^*}$, and mass $m_{K^*}^*$ of the vector meson.

However, for the strange axial vector meson channel, the spectral density $R^{K_1}_{\rm phen}(s)$ will have a contribution from the pseudoscalar $K$ meson as well as from the axial vector $K_1$ meson resonance and we parameterize the spectral density for strange axial vector $K_1$ meson as,
\begin{equation}
R^{K_1}_{\rm phen}(s)=f_K^2\,\delta({s-m_{K}^2}) + F_{K_1}\,\delta(s-m_{K_1}^2)+d_{K_1}c_0^{K_1}\,\Theta(s-s_0^{K_1})
\label{simple_ansatz_K1}
\end{equation}
with $f_K$ and $m_K$ being the kaon decay constant and kaon mass respectively. A similar parameterization scheme have been used for non-strange axial vector $A_1$ meson which gets additional contribution to the spectral density due to pseudoscalar pion ($\pi$) \cite{Leupold_PRC64_015202_2001, Shifman_NPB147_448_1979, Parui_arXiv_2209_02455}. The FESRs for the strange axial vector meson in the nuclear medium are given as,
\begin{equation}
 F_{K_1}^*=d_{K_1}(c_0^{K_1} s_0^{*{K_1}}+c_1^{K_1})-f_K^2
 \label{FESR_K1_med1}
\end{equation}
\begin{equation}
F_{K_1}^* \,m_{K_1}^{*2} = d_{K_1}\Big(\frac{(s_0^{*{K_1}})^2 c_0^{K_1}}{2}-c_2^{*{K_1}}\Big)-f_K^2 m_K^2
\label{FESR_K1_med2}
\end{equation}
\begin{equation}
 F_{K_1}^* \,m_{K_1}^{*4} = d_{K_1}\Big(\frac{(s_0^{*{K_1}})^3 c_0^{K_1}}{3}+c_3^{*{K_1}}\Big) -f_K^2 m_K^4 
 \label{FESR_K1_med3}
\end{equation}
The above FESRs are solved simultaneously to find the in-medium masses for the axial vector $K_1$ mesons.

\subsection{Effects of strong magnetic fields on $K^*$ meson masses}
(1) Whenever a charged particle moves in an external magnetic field, the Lorentz force comes into play and the particle's momenta perpendicular to the direction of magnetic field, are discretized to certain levels characterized by an integral label $(n)$ called Landau levels; while the particle's momenta in the direction of magnetic field remains unaffected. Therefore the energy levels of charged pseudoscalar (spin-0) and vector (spin-1) mesons, in the presence of magnetic field, are discretized to different Landau levels \cite{Gubler_PRD93_054026} and are given by, 
\begin{equation}
m_P(eB)= \sqrt{m_P^2 + (2n+1)|eB|+p_z^2}
\label{Landau_pseudo_1}
\end{equation}
\begin{equation}
m_V(eB) = \sqrt{m_V^2 + (2n+1)|eB|+p_z^2+gS_z|eB|}
\label{Landau_vector_1}
\end{equation}
where $m_P$ and $m_V$ are the masses in the absence of magnetic field. The integer `$n$' specifies the Landau levels, $p_z$ is the continuous momentum in the z-direction, g is the Lande g-factor, and $S_z$ is the spin quantum number along the direction of magnetic field. The internal structure of the mesons is not considered while writing the above expressions \cite{Chernodub_LNP871_143_2013}. In the present discussion, we will consider only the lowest Landau level (LLL), $n=0$, contribution at zero momentum in the z-direction ($p_z=0$). The contribution from the higher Landau levels ($n\geq 1$) is more important in the weak magnetic field case as these will be very close to the lowest Landau level and hence can not be treated as the continuum part. The three polarization states ($S_z= +1,0,-1$) of charged vector $K^{*+}$ meson have different Landau contributions to the masses, and pseudoscalar $K$ meson mass also gets modified, which are given by, 
\begin{equation}
m_P(eB)= \sqrt{m_P^2 + eB}
\label{Landau_pseudo_2}
\end{equation}
\begin{equation}
m_{V^{\perp}_{+1}}(eB) = \sqrt{m_V^2 + 3\,eB},\; m_{V^{||}}(eB) = \sqrt{m_V^2 + eB},\;m_{V^{\perp}_{-1}}(eB)= \sqrt{m_V^2 - eB}
\label{Landau_vector_2}
\end{equation}

(2) The presence of external magnetic field can induce mixing among some spin states due to breaking of a part of spatial rotation symmetry. Because to this, only the mesonic spin states oriented along the direction of the magnetic field can persist as a good quantum number and there is a mixing between the pseudoscalar meson \big(($S,S_z) = (0,0$)\big) and longitudinal part \big(($S,S_z) = (1,0$)\big) of vector meson. The PV mixing effect can be incorporated through an effective interaction vertex, ensuring the Lorentz invariance, in the Lagrangian as \cite{Cho_PRD91_045025_2015, AM_PRC102_045204_2020, AM_IJMPE30_2150064_2021},
\begin{equation}
\mathcal{L}_{\gamma PV}= \frac{g_{PV}}{m_{\rm avg}}e \Tilde{F}_{\mu\nu}(\partial^\mu P)V^\nu
\label{L_PV_mixing}
\end{equation}
where $m_{\rm avg}$ is the average of masses of pseudoscalar ($m_P$) and longitudinal part of vector meson ($m_{V^{||}})$. Here $P$ and $V^\mu$ are the pseudoscalar and vector meson fields and $\Tilde{F}_{\mu\nu}$ is the dual field strength tensor of QCD. Here, we considered the magnetic field in z-direction so that the only non-zero components of dual field strength tensor are $\Tilde{F}_{03} = -\Tilde{F}_{30} = B$ and the mesons are considered to be at rest. We fit the dimensionless coupling parameter $g_{PV}$ from the experimentally observed radiative decay width $\Gamma(V\rightarrow P\gamma) $\cite{Gubler_PRD93_054026,AM_PRC102_045204_2020,AM_IJMPE30_2150064_2021}. From the equations of motion, obtained from the effective Lagrangian containing the kinetic and interaction term, we find that only the longitudinal part $V^{||}$ is mixed with pseudoscalar P and there is no mixing for the transverse parts $V^{\perp}$ of vector meson. The energy eigenvalues for the physical state are given by \cite{Gubler_PRD93_054026,Cho_PRD91_045025_2015,AM_PRC102_045204_2020,AM_IJMPE30_2150064_2021}, 
\begin{equation}
m_{V^{||},P}^{PV}=\sqrt{\frac{1}{2}\Bigg(M_+^2 + \frac{c_{PV}^2}{m_{\rm avg}^2}\pm\sqrt{M_-^4 + \frac{2c_{PV}^2 M_+^2}{m_{\rm avg}^2}+\frac{c_{PV}^4}{m_{\rm avg}^4}}\Bigg)}
\label{M_PV_mixing}
\end{equation}
where $M_+^2=m_V^2 + m_P^2$, $M_-^2=m_V^2 - m_P^2$, and $c_{PV}=g_{PV}\,eB$. The effects of PV mixing, on charged meson masses, can be studied separately with and without Landau contributions on the masses, while the neutral mesons do not have contribution due to Landau quantization. 

(3) Furthermore, the PV mixing effects can also be studied through the spin-magnetic field interaction ($-\mu_i.B$) term. The external magnetic field with quantum numbers ($J^P = 1^+$) can induce mixing between a pseudoscalar ($J^P = 0^-$) and vector meson ($J^P = 1^-$). The spin singlet state $\big|00\big>$ is mixed with longitudinal component $\big|10\big>$ of spin triplet state,
\begin{equation}
\big(-\mu_i.B\big)\big|00\big>=-\frac{gB}{4}\Big(\frac{q_1}{m_1}-\frac{q_2}{m_2}\Big)\big|10\big>
\label{spin_mix_longitudinal}
\end{equation}
where the magnetic moment of the $i$th particle (with $i=1,2$ corresponding to quark, antiquark in this work) is $\mu_i=gq_i\sigma_i/4m_i$, where $\sigma_i$ are the Pauli spin matrices. The Lande factor $g$ is taken to be 2 and ($q_i, m_i$) are the charges and constituent masses of quark/antiquark respectively. Since the $\big|10\big>$ and $\big|00\big>$ states are not pure eigenstates of interaction Hamiltonian term \big($\mathcal{H}_s= - \mu_i.B$\big), we will consider a two dimensional eigensystem, for the $\big|10\big>$ and $\big|00\big>$ states, to determine the spin mixing effect \cite{Alford_PRD88_105017_2013}. Then effective physical mass eigenvalues are given by,
\begin{equation}
m_{V^{||},P}^{\rm eff}= m_{V^{||},P}\pm\frac{\Delta E}{2}\big(\sqrt{1+\chi_{sB}^2}-1\big)
\label{mass_spin_mix_longitudinal}
\end{equation}
where $\Delta E= m_{V^{||}} - m_P$, and $\chi_{sB}=\big(\frac{2g|eB|}{\Delta E}\big)\Big(\frac{q_1}{m_1}-\frac{q_2}{m_2}$\Big). For the charged meson, the effects of Landau levels are also considered here in $m_{V^{||},P}$. The transverse polarized states, $\big|1 +1\big>$ and $\big|1 -1\big>$, do not get mixed with any other states \cite{Yoshida_PRD94_074043_2016,Machado_PRD88_034009_2013,Iwasaki_EPJA57_222_2021}. The energy eigenvalue equation for transverse polarized states is given by, 
\begin{equation}
\big(-\mu_i.B\big)\big|1 \pm 1\big>=\mp\frac{gB}{4}\Big(\frac{q_1}{m_1}+\frac{q_2}{m_2}\Big)\big|1 \pm 1\big>
\label{spin_mix_transverse}
\end{equation}
This contribution vanishes for meson states, like charmonium and bottomonium, which have same quark-antiquark as constituents due to $q_1 = -q_2$ and $m_1 = m_2$. The quark masses taken here are the constituent quark masses.

\section{Decay Width of $K^{*} \rightarrow K \pi$ and $K_1 \rightarrow K^{*} \pi$ in a Phenomenological Model}
\label{self_energy_section}
(1) We now discuss the medium modifications of masses and decay widths of $K^*$ meson from the self energy at one-loop level. The interaction term for a vector meson V decaying to two pseudoscalar mesons $P_1$ and $P_2$ can be written as \cite{Klingl_ZPA356_193_1996},
 \begin{equation}
\mathcal{L}_{V\bar{P_1}P_2}=igV^\mu[\bar{P_1}(\partial_\mu P_2)-(\partial_\mu \bar{P_1})P_2]
\label{L_V_P1_P2}
\end{equation}
The parameter `$g$' is known as the coupling strength of the decay channel. This interaction term of the effective Lagrangian generates a hadronic current, which couples with the $K^*$ meson field to produce the self energy, which is given by,
\begin{equation}
    \Pi^{\mu \nu}(p)=-ig_{K^*}^2\int\frac{d^4q}{(2\pi)^4}\frac{1}{q^2-m_K^2+i\epsilon}\Big[\frac{q^\mu q^\nu}{(q-p)^2-m_\pi^2+i\epsilon}\Big]
    \label{self_energy_eq}
\end{equation}
where $p$ is the $4$ -momenta carried by vector $K^*$ meson and the $4$ -momentas of intermediate $K$ and $\pi$ are $q$ and ($q-p$) respectively, and we have integrated over the internal loop momenta. 

\begin{figure}[htp]
\begin{center}
\includegraphics[width=14cm, height=6cm]{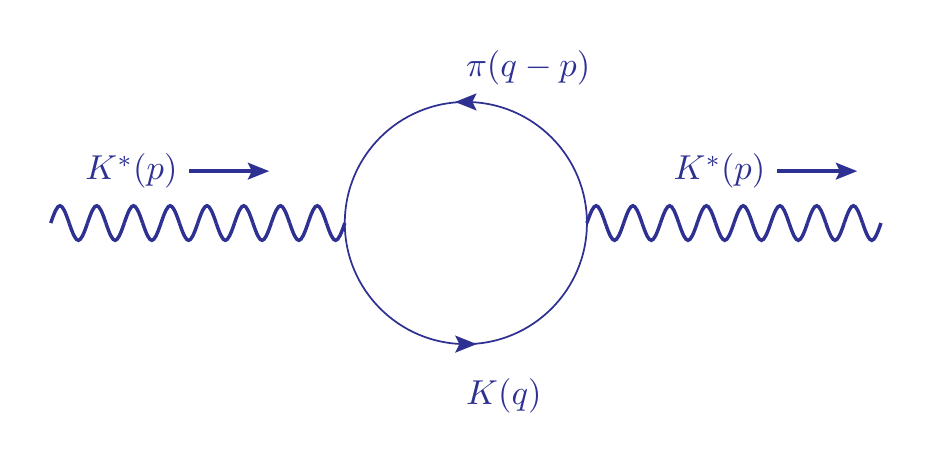}
\caption{ $K^*$ meson self energy diagram at one loop level}
\label{self_energy_diagram_K*}
\end{center}
\end{figure}

Writing  $\Pi_{K^*}(p^2)=\frac{1}{3} g_{\mu\nu} \Pi^{\mu\nu}(p)$ and $\mathring{m}_{K^*}$ as the bare $K^*$ meson mass, the physical mass can be written as \cite{Cobos-Martinez_PLB771_113_2017},
\begin{equation}
m_{K^*}^2 = \mathring{m}_{K^*}^2 + Re\,\Pi_{K^*}(p^2=m_{K^*}^2)
 \label{mass_bare_mass_relation}
\end{equation}

The decay width, at resonance, is related to the imaginary part of the propagator as,
\begin{equation}
\Gamma(K^*\rightarrow K\pi)=-Im\,\Pi_{K^*}(p^2=m_{K^*}^2)/m_{K^*}
 \label{DW_self_energy}
\end{equation}
The imaginary part of the propagator is calculated by the standard Cutkosky rule \cite{Itzykson_QFT_1980} and  is given by, 
\begin{equation}
Im\,\Pi_{K^*}(p^2)=-\frac{g_{K^*}^2 m_{K^*}^2}{64\pi}\Big[1-\frac{(m_K-m_\pi)^2}{m_{K^*}^2}\Big]^{3/2}\Big[1-\frac{(m_K+m_\pi)^2}{m_{K^*}^2}\Big]^{3/2}
 \label{imaginary_part_self_energy}
\end{equation}
The scalar part of the in-medium self-energy, in the rest frame of $K^*$, can be written as,
 \begin{equation}
 i\Pi_{K^*}(p)=-\frac{1}{3}g_{K^*}^2 \int \frac{d^4q}{(2\pi)^4}\Vec{q^2}D_K(q)D_{\pi}(q-p) 
 \label{scalar_part_self_energy}
\end{equation}

 where $D_K(q)=(q^2-m^2_K+i\epsilon)^{-1}$ and $D_{\pi}(q)=\big((q-p)^2-m^2_\pi+i\epsilon\big)^{-1}$ are the kaon and pion propagators, respectively. The real part of the self energy is written as, 
\begin{equation}
 Re\,\Pi_{K^*}(p)=-\frac{1}{3}g_{K^*}^2 \mathcal{P}\int \frac{d^3q}{(2\pi)^3}\Vec{q^2}\frac{(E_K+E_\pi)}{(2E_K E_\pi)\big((E_K+E_\pi)^2-m_{K^*}^2\big)}
 \label{real_part_self_energy}
\end{equation}
Here $\mathcal{P}$ denotes the principal value of the integral and the energies are given by $E_K=\sqrt{\Vec{q^2}+m_K^2}$ and $E_\pi=\sqrt{\Vec{q^2}+m_\pi^2}$. The integral (\ref{real_part_self_energy}) is divergent and it needs to be regularized to avoid singularities. We use a phenomenological form factor approach with a cutoff parameter $\Lambda_c$ \cite{Leinweber_PRD64_094502_2001,Krein_PLB697_136_2011}, and the integral then becomes, 
\begin{align}
Re\,\Pi_{K^*}(p)=-\frac{1}{3}g_{K^*}^2 \mathcal{P}\int_{0}^{\Lambda_c} \frac{d^3q}{(2\pi)^3}\Vec{q^2}&\Big(\frac{\Lambda_c^2 + m_{K^*}^2}{\Lambda_c^2 + 4E_K^2}\Big)^2\Big(\frac{\Lambda_c^2 + m_{K^*}^2}{\Lambda_c^2 + 4E_\pi^2}\Big)^2\nonumber\\&\frac{(E_K+E_\pi)}{(2E_K E_\pi)\big((E_K+E_\pi)^2-m_{K^*}^2\big)}
 \label{real_part_cutoff_self_energy}
\end{align}
where the quantities, called vertex form factors, are given by,
\begin{equation}
 u_{K,\pi}(q^2)=\Bigg(\frac{\Lambda_c^2 + m_{K^*}^2}{\Lambda_c^2 + 4E_{K,\pi}^2}\Bigg)^2
 \label{form_factors}
\end{equation}
The coupling parameter $g_{K^*}$ is determined separately, from the empirical decay width of the $K^*$ meson in vacuum, for each particular value of cut-off parameter $\Lambda_c$. Then we fix the bare mass of the $K^*$ meson by the relation (\ref{mass_bare_mass_relation}), from the vacuum mass of $K^*$ meson. In a simple picture, the cutoff parameter is related to the overlap region of the parent and daughter particles at the vertex and depends on the size of their wave functions as done in \cite{Krein_PLB697_136_2011, Lee_PRC67_038202_2003}. So we also calculate the vertex form factors using the ${}^3 P_0$ model for quark-antiquark pair creation with Gaussian wave functions for the mesons. Then the root mean square radii from the two form factors is compared to get an estimate of the cutoff mass $\Lambda_c$ as done in Ref. \cite{Krein_PLB697_136_2011} for the $J/\psi$ meson.  A similar work has been done for the $\phi$ meson in \cite{Cobos-Martinez_PLB771_113_2017}. To include the uncertainty in the estimated value, we take the range of $\Lambda_c$ from $1$ to $4$ GeV. 

(2) Further, we also use a phenomenological approach for the study of $K_1 \rightarrow K^* \pi$ decay. The interaction Lagrangian is constructed from the anti-symmetric tensor fields for the axial-vector and vector meson. The $SU(3)$ matrices associated with the tensor fields for the two axial vector nonets $a_1$ and $b_1$, and vector meson nonet are, respectively given as

\vspace{0.4 cm}
$A_{\mu\nu}$ =
${\begin{pmatrix}
\frac{1}{\sqrt{2}}a_1^0 + \frac{1}{\sqrt{2}}f_1(1285)& a_1^+&K_{1A}^+ \\
a_1^-&-\frac{1}{\sqrt{2}}a_1^0 + \frac{1}{\sqrt{2}}f_1(1285)&K_{1A}^0 \\
K_{1A}^-&\bar{K_{1A}^0}&f_1(1420) 
\end{pmatrix}}_{\mu\nu}$

$B_{\mu\nu}$ =
${\begin{pmatrix}
\frac{1}{\sqrt{2}}b_1^0 + \frac{1}{\sqrt{2}}h_1(1170)& b_1^+&K_{1B}^+\\
b_1^-&-\frac{1}{\sqrt{2}}b_1^0 + \frac{1}{\sqrt{2}}h_1(1170)&K_{1B}^0\\
K_{1B}^-&\bar{K_{1B}^0}&h_1(1380)
\end{pmatrix}}_{\mu\nu}$
\vspace{0.4 cm}

$V_{\mu\nu}$ =
${\begin{pmatrix}
\frac{1}{\sqrt{2}}\rho^0 + \frac{1}{\sqrt{2}}\omega^0 & \rho^+ &K^{*+}\\
\rho^-&-\frac{1}{\sqrt{2}}\rho^0 + \frac{1}{\sqrt{2}}\omega^0 &K^{*0}\\ K^{*-}&\bar{K^{*0}}&\phi
\end{pmatrix}}_{\mu\nu}$
\vspace{0.4 cm}

The mesons corresponding to two nonets, $a_1$ and $b_1$ above, corresponds to $J^{PC} = 1^{++}$ and $1^{+-}$ states. The interaction vertex, ensuring the $SU(3)$ invariance, charge conjugation (C), and parity (P) conservation, for the decay of an axial vector meson A (and B) to a vector meson V and a pseudoscalar meson P is written as \cite{Roca_PRD70_094006_2004},
\begin{equation}
 \mathcal{L}_{AVP}= i\Tilde{F}\big<A_{\mu\nu}[V^{\mu\nu},P]\big>
 \label{L_AVP}
\end{equation}
\begin{equation}
\mathcal{L}_{BVP}= \Tilde{D}\big<B_{\mu\nu} \{V^{\mu\nu},P\}\big>
\label{L_BVP}
\end{equation}
where $P$ is the usual $SU(3)$ matrix for the pseudoscalar meson nonet considering the standard $\eta-\eta^{'}$ mixing. The free parameters, $\Tilde{F}$ and $\Tilde{D}$, are fitted globally from the available data of various decays and branching ratios of the members of $a_1$ and $b_1$ nonets in Ref. \cite{Roca_PRD70_094006_2004} and the mixing angle is taken to be $\theta_{K_1}$ $\sim$ $62^{\circ}$. The symbol $\big<...\big>$ represents the $SU(3)$ trace and factor $`i$' ensures that the Lagrangian is hermitian. The decay width for the $K_1\rightarrow K^* \pi$ decay calculated within the phenomenological approach is given as,
\begin{equation}
 \Gamma_{K_1\rightarrow K^* \pi}= \frac{q}{8\pi M_{K_1}^2}{\mathcal{|M|}^2}
 \label{DW_K1_phen1}
\end{equation}
where the momentum $q$ of the final state particles, in the rest frame of the parent particle $K_1$, is given by,
\begin{equation}
 q(M_{K_1},M_{K^*},M_\pi) = \frac{1}{2M_{K_1}}\sqrt{\big(M_{K_1}^2-(M_{K^*}+M_\pi)^2\big)\big(M_{K_1}^2-(M_{K^*}-M_\pi)^2\big)}
 \label{momentum_phen}
\end{equation}
and the matrix element is given by,
\begin{equation}
 \mathcal{M} = \frac{-2\lambda_{AVP}}{M_{K_1}M_{K^*}}(p'.p\,\epsilon'.\epsilon - \epsilon'.p\,\epsilon.p')
  \label{matrix_element_phen}
\end{equation}
with $p',\epsilon'$ and $p,\epsilon$ being the momentum and polarization vector of the axial-vector $K_1$ and vector $K^*$ meson, respectively. The decay width then becomes,
\begin{equation}
\Gamma_{K_1\rightarrow K^* \pi} = \frac{|\lambda_{AVP}|^2}{2\pi M_{K_1}^2}q\Big(1+\frac{2}{3}\frac{q^2}{M_{K^*}^2}\Big)
 \label{DW_K1_phen2}
\end{equation}
 The coefficients $\lambda_{AVP}$ of the interaction vertex are given as $\pm\frac{1}{\sqrt{2}}({\rm cos}\,\theta \,\Tilde{D}+{\rm sin}\,\theta \,\Tilde{F})$, $\pm({\rm cos}\,\theta \,\Tilde{D}+{\rm sin}\,\theta \,\Tilde{F})$, $\pm({\rm cos}\,\theta \,\Tilde{D}+{\rm sin}\,\theta \,\Tilde{F})$, and $+\frac{1}{\sqrt{2}}({\rm cos}\,\theta \,\Tilde{D}+{\rm sin}\,\theta \,\Tilde{F})$ for the $K_1^{\pm}\rightarrow K^{*\pm}\pi^0$, $K_1^{\pm}\rightarrow K^{*0}\pi^\pm$, $K_1^{0}\rightarrow K^{*\pm}\pi^\mp$, and $K_1^{0}\rightarrow K^{*0}\pi^0$ decays respectively, with $\theta \equiv \theta_{K_1}$.

\section{Decay Widths of $K^* \rightarrow K \pi$ and $K_1 \rightarrow K^* \pi$ within the ${}^3 P_0$ model}
\label{DW_section}
\hspace{0.5cm} (1) First, we study the decay width of vector $K^*$ meson to two pseudoscalar mesons which are pion and kaon, using the ${}^3 P_0$ model \cite{Barnes_PRD55_4157_1997, Friman_PLB548_153_2002}. This model was first introduced by  L. Micu to calculate the decay rates of various meson resonances \cite{Micu_NPB10_521_1969} and then extended further by A. Le Yaouanc and others to explain strong decay amplitudes of mesons and baryons \cite{Yaouanc_PRD8_2223_1973}. In this model, a light quark-antiquark pair is assumed to be produced in the ${}^3 P_0$ state having vacuum quantum numbers $J^{PC} = 0^{++}$. The quark (antiquark) of this produced pair combines with the antiquark (quark) of the parent meson, which is assumed to be at rest initially, to give the final state mesons. The matrix element for the general decay $A\rightarrow BC$ in the ${}^3 P_0$ model is given as, 
\begin{equation}
\mathcal{M}_{A\rightarrow BC}= \big<A\big|\gamma\,[\bar{q_s}q_s]^{{}^3 P_0}\big|BC\big>
\label{general_M_3P0}
\end{equation}
where $\gamma$ is the coupling strength which is related to the probability of production of a quark antiquark pair in the ${}^3 P_0$ state. The wave functions for the produced $\bar{q}q$ pair are chosen to be simple harmonic oscillator (SHO) wave functions to calculate the strong decay amplitude. In the present case of a vector $K^*$ meson decaying to two pseudoscalar mesons, the matrix element for the $1({}^3S_1)\rightarrow1({}^1S_0) + 1({}^1S_0)$ channel with proper flavor factor $I_f$ is given by,
\begin{equation}
\mathcal{M}_{K^*\rightarrow K\pi}= \frac{\gamma_{K^*}}{\pi^{1/4}\beta_{\rm avg}^{1/2}}\bigg[\frac{-2^4}{3^{1/2}}\frac{r^{3/2}(1+r^2)x}{(1+2r^2)^{5/2}}\bigg] {\rm exp}\Big(\frac{-x^2}{4(1+2r^2)}\Big)I_f
\label{Matrix_element_3P0_K*}
\end{equation}
where the ratio $r=\frac{\beta_{K^*}}{\beta_{\rm avg}}$, with $\beta_{K^*}$ being the harmonic oscillator potential strength for the $K^*$ meson and $\beta_{\rm avg}$ as the average of harmonic oscillator potential strength of two daughter mesons. Taking $r\neq 1$ i.e. $\beta_{K^*}\neq \beta_{\rm avg}$ allows to account for the different sizes of daughter and parent mesons. The factor $I_f$ gives the flavor weight factor contributions from the two Feynman decay diagrams, called $d_1$ and $d_2$ diagrams, of the parent meson \cite{Barnes_PRD55_4157_1997} and is taken to be $1/(2\sqrt{2})$ for the above decay. The quantity `$x$' is the scaled momentum, given by $p_K/\beta_{\rm avg}$, carried by the daughter meson. The decay width is then given by, 
\begin{equation}
\Gamma({K^*}\rightarrow K\pi)= 2\pi \frac{p_K E_K E_\pi}{M_{K^*}}\sum_{LS}\Big|\mathcal{M}_{LS}\Big|^2
\label{DW_3P0_K*}
\end{equation}
where $p_K$ is the 3-momentum, given by, 
\begin{equation}
p_K= \sqrt{\frac{M_{K^*}^2}{4}-\frac{M_K^2 +M_\pi^2}{2}+\Big(\frac{M_K^2-M_\pi^2}{2M_{K^*}}\Big)^2}
\label{momentum_3P0}
\end{equation}
 and the energies are given by $E_K=\sqrt{p_K^2+m_K^2}$ and $E_\pi=\sqrt{p_\pi^2+m_\pi^2}$. As the decaying meson is assumed to be at rest, momentum conservation gives $p_K=p_\pi$. The masses $M_{K^*}$ and $M_K$ are the in-medium masses, from which we calculate the in-medium decay width. In this work, we do not take into account the medium modification of the pion mass.

(2) The axial vector $K_1(1270)$ meson is not a pure $1{}^3P_1$ ($J^{PC}=1^{++}$) or $1{}^1P_1$ ($J^{PC}=1^{+-}$) state, but the physically observed $K_1(1270)$ and $K_1(1400)$ are a mixture of the two non mass eigenstates, $\big|K_{1A}\big>$ and $\big|K_{1B}\big>$, of the two strange axial vector nonets $1{}^3P_1$ and $1{}^1P_1$, respectively \cite{Tayduganov_PRD85_074011_2012,Suzuki_PRD47_1252_1993},
\begin{equation}
\big|K_1(1270)\big> = {\rm sin}\,\theta_{K_1}\big|K_{1A}\big>+ {\rm cos}\,\theta_{K_1} \big|K_{1B}\big> 
\label{K1_1270}
\end{equation}
\begin{equation}
\big|K_1(1400)\big> = {\rm cos}\,\theta_{K_1}\big|K_{1A}\big>- {\rm sin}\,\theta_{K_1} \big|K_{1B}\big> 
\label{K1_1400}
\end{equation}
where $\theta_{K_1}$ is the mixing angle. The matrix element for the axial vector meson decay channel ($K_1\rightarrow K^*\pi$) is given as, 
\begin{equation}
\mathcal{M}_{LS}= \frac{\gamma_{K_1}}{\pi^{1/4}\beta_{\rm avg}^{1/2}}\mathcal{P}_{LS}(x,r)\,{\rm exp}\Big(\frac{-x^2}{4(1+2r^2)}\Big)I_f
\label{Matrix_element_3P0_K1}
\end{equation}
where the polynomials for various decays are given as \cite{Barnes_PRD55_4157_1997,Friman_PLB548_153_2002}, 
\begin{equation}
\mathcal{P}_{01}^{\big(1{}^3P_1\rightarrow1{}^3S_1+1{}^1S_0\big)}= \Bigg(2^5\Big(\frac{r}{1+2r^2}\Big)^{5/2}\Big(1-\frac{1+r^2}{3(1+2r^2)}x^2\Big)\Bigg)
\label{polynomial_K1_1}
\end{equation}

\begin{equation}
\mathcal{P}_{21}^{\big(1{}^3P_1\rightarrow1{}^3S_1+1{}^1S_0\big)}= -\sqrt{\frac{5}{6}}\Bigg(\frac{1}{\sqrt{15}}\frac{r^{5/2}2^5(1+r^2)}{(1+2r^2)^{7/2}}x^2\Bigg)
\label{polynomial_K1_2}
\end{equation}

\begin{equation}
\mathcal{P}_{01}^{\big(1{}^1P_1\rightarrow1{}^3S_1+1{}^1S_0\big)}= -\frac{1}{\sqrt{2}}\Bigg(2^5\Big(\frac{r}{1+2r^2}\Big)^{5/2}\Big(1-\frac{1+r^2}{3(1+2r^2)}x^2\Big)\Bigg)
\label{polynomial_K1_3}
\end{equation}

\begin{equation}
\mathcal{P}_{21}^{\big(1{}^1P_1\rightarrow1{}^3S_1+1{}^1S_0\big)}= -\sqrt{\frac{5}{3}}\Bigg(\frac{1}{\sqrt{15}}\frac{r^{5/2}2^5(1+r^2)}{(1+2r^2)^{7/2}}x^2\Bigg)
\label{polynomial_K1_4}
\end{equation}

with $r=\frac{\beta_{K_1}}{\beta_{\rm avg}}$, and $\beta_{\rm avg}$ being the average of harmonic oscillator potential strength of the daughter $K^*$ and $\pi$ mesons. The flavor factor $I_f$ is scaled together with the coupling constant $\gamma_{K_1}$ in this work. The decay width for $K_1\rightarrow K^*\pi$ decay is then evaluated from equation (\ref{DW_3P0_K*}) after changing the corresponding variables for this decay.

\section{Results and Discussion}
\label{result_section}
Now we discuss the medium modifications of the $K^*$ and $K_1$ meson masses and decay widths in nuclear matter including the effects of isospin asymmetry, and the effects of the magnetic field on the vector $K^*$ meson will also be discussed. Isospin asymmetry corresponds to the fact that the number of neutrons in heavy ion collision experiments is more than the number of protons inside the colliding heavy ions. Isospin asymmetry factor ($\eta$) gives the amount of isospin asymmetry in the medium. For example, $\eta = 0.5$ corresponds to a medium consisting of neutrons only. We investigate the in-medium masses and decay widths at different values ($\rho_0$, $2 \rho_0$, and $ 4\rho_0 $) of medium density where nuclear matter saturation density, $\rho_0$ = $0.15 \, {\rm fm}^{-3}$.

First, the medium modifications of scalar fields ($\sigma$, $\zeta$, $\delta$, and $\chi$) are calculated using a chiral effective model by solving the coupled equations of motion for these fields. Then these modifications are included in scalar gluon condensates, $\big<\frac{\alpha_s}{\pi}G^a_{\mu \nu}G^{a\mu\nu}\big>$ and light quark condensates \Big($\big<\bar{u}u\big>,\big<\bar{d}d\big>, \big<\bar{s}s\big>$\Big). Finally, we calculate the in-medium masses using QCD sum rule (QCDSR) approach from the medium modified scalar quark and gluon condensates, and then in-medium decay widths are calculated from the in-medium masses, using the ${}^3 P_0$ model. The in-medium masses and decay widths for the $K^*$ meson are also calculated from the in-medium self energy of the $K^*$ meson at the one-loop level as discussed in section \ref{self_energy_section}. 

The values of constant parameters in sections \ref{SU3_model_section} and \ref{QCDSR_section}, taken in this work, are given as $\alpha_s(Q^2= 1$ GeV$^2)=0.3551$, $d=0.064018$, $m_\pi$=139 MeV, pion decay constant $f_\pi=93.3$ MeV, $m_K$= 498 MeV, kaon decay constant $f_K= 122.143$ MeV. The vacuum values of the scalar fields are $\sigma_0=-93.3$ MeV, $\zeta_0=-106.6$ MeV, and $\chi_0$= 409.77 MeV. The vacuum mass values of $K^{*+}$, $K^{*0}$, and $K_1$ mesons are taken to be 891.67, 895.55, and 1253 MeV respectively \cite{AM_IJMPE30_2150014_2021, Zyla_PDG_2020}. The current quark masses, used in QCD sum rule approach, for the light quarks are taken as $m_u= 4$ MeV, $m_d= 7$ MeV, and $m_s= 150 $ MeV. The isospin asymmetry parameter, in the medium consisting of nucleons only, is given by $\eta= \frac{\rho_n-\rho_p}{2\rho_B}$, where ($\rho_n, \rho_p$) are the vector number density of (neutrons, protons) respectively with $\rho_B$ as total baryon density. The coefficient $\kappa_n$ for $K^*$ and $K_1$ mesons is found from the vacuum FESRs with the values of vacuum masses and values of the scalar quark and gluon condensates. The obtained values are 0.73586, 2.30955, 22.53633, and 17.77003 for the $K^{*+}$, $K^{*0}$, $K_1^+$, and $K_1^0$ mesons, respectively.

\begin{figure}[htp]
\includegraphics[width=\linewidth]{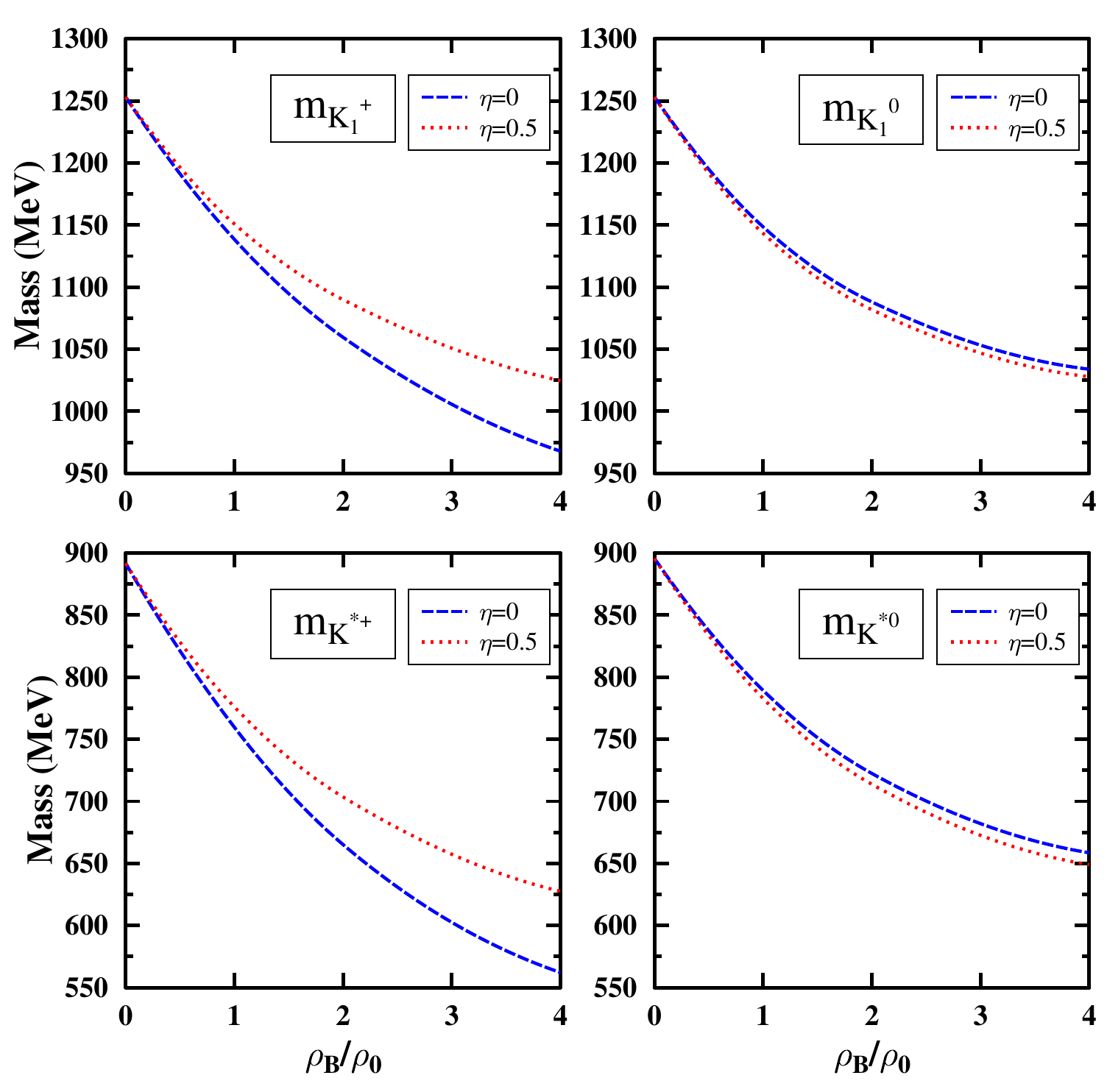}
\caption{Masses of vector $K^*$ meson and axial vector $K_1$ meson plotted as a function of nuclear medium density ($\rho_B$) in terms of nuclear matter saturation density $\rho_0$ for various values of isospin asymmetry parameter $\eta$, within QCD sum rule approach at zero magnetic field.}
\label{masses_vector_axial_vector_QCDSR}
\end{figure}

\subsection{In-medium Masses using QCD Sum Rule Approach}
In figure (\ref{masses_vector_axial_vector_QCDSR}), the masses of $K^{*}$ and $K_1$ mesons are plotted as a function of nuclear medium density in terms of nuclear matter saturation density (${\rho_0}$) for various isospin asymmetry parameters $(\eta)$, calculated within the QCD sum rule approach. The mass values are observed to decrease monotonically as the medium density increases. This is because of the fact that the chiral condensates $\big<\bar{q}q\big>$ and the gluon condensates $\big<\frac{\alpha_s}{\pi}G^a_{\mu\nu}G^{a\mu\nu}\big>$ decrease in magnitude as a function of medium density. The non-strange quark condensates \Big($\big<\bar{u}u\big>$, $\big<\bar{d}d\big>$\Big) decrease more in nuclear matter as compared to strange quark condensate $\big<\bar{s}s\big>$ and gluon condensates \cite{AM_PRC100_015207_2019}. There is observed to be a smaller mass drop in isospin asymmetric nuclear matter ($\eta\neq 0$) as compared to isospin symmetric matter ($\eta = 0$) for the charged $K^{*+}$ and $K_1^+$ meson. The mass modifications for the neutral $K^{*0}$ and $K_1^0$ mesons are observed to be smaller as compared to the corresponding charged mesons and the mass drop is slightly higher in $\eta\neq0$ matter, for the neutral mesons, as compared to isospin symmetric matter. Also, there is observed to be a sharper mass drop for smaller values of nuclear medium density up to around 2$\rho_0$. As nuclear medium density increases, the drop in strange quark condensates decreases even more which results in smaller changes in $K^*$ and $K_1$ meson mass at higher densities. However, the isospin asymmetry effects are observed to be larger at higher medium density. The masses for these open strange mesons are given in table I and II as calculated within the framework of QCD sum rule approach.   
\begin{center}
\textbf{Table I: Masses of Vector $K^*$ Meson (QCD Sum Rule Approach) }\\
\vspace{0.3cm}
\begin{tabular}{|p{2.5cm}||p{2.5cm}|p{2.5cm}||p{2.5cm}|p{2.5cm}|  }
 \hline
&\multicolumn{2}{|c||}{mass of $K^{*+}$ (MeV)}&\multicolumn{2}{|c|}{mass of $K^{*0}$ (MeV)} \\
 \hline
 \hline
 density & $\eta=0$ & $\eta =0.5$ & $\eta=0$ & $\eta =0.5$\\
 \hline
 \hline
 $\rho_0$   & 759.67  &776.05 &   789.28 & 782.79\\
 \hline
 2$\rho_0$&   665.18  & 703.53  & 722.43 & 713.72\\
 \hline
 4$\rho_0$ & 562.15 & 627.55 &  658.54 & 649.02\\
\hline
\end{tabular}
\end{center}
\vspace{0.3cm}
\begin{center}
\textbf{Table II: Masses of Axial Vector $K_1$ Meson (QCD Sum Rule Approach) }\\
\vspace{0.3cm}
\begin{tabular}{|p{2.5cm}||p{2.5cm}|p{2.5cm}||p{2.5cm}|p{2.5cm}|  }
 \hline
&\multicolumn{2}{|c||}{mass of $K_1^+$ (MeV)}&\multicolumn{2}{|c|}{mass of $K_1^0$ (MeV)} \\
 \hline
 \hline
 density & $\eta=0$ & $\eta =0.5$ & $\eta=0$ & $\eta =0.5$\\
 \hline
 \hline
 $\rho_0$   & 1138.45  &1150.91 &   1148.44 & 1143.18\\
 \hline
 2$\rho_0$&   1059.59  & 1089.87  & 1088.04 & 1081.76\\
 \hline
 4$\rho_0$ & 967.85 & 1024.71 &  1033.85 & 1027.84\\
\hline
\end{tabular}
\end{center}

\begin{figure}[hbt!]
\includegraphics[width=\linewidth]{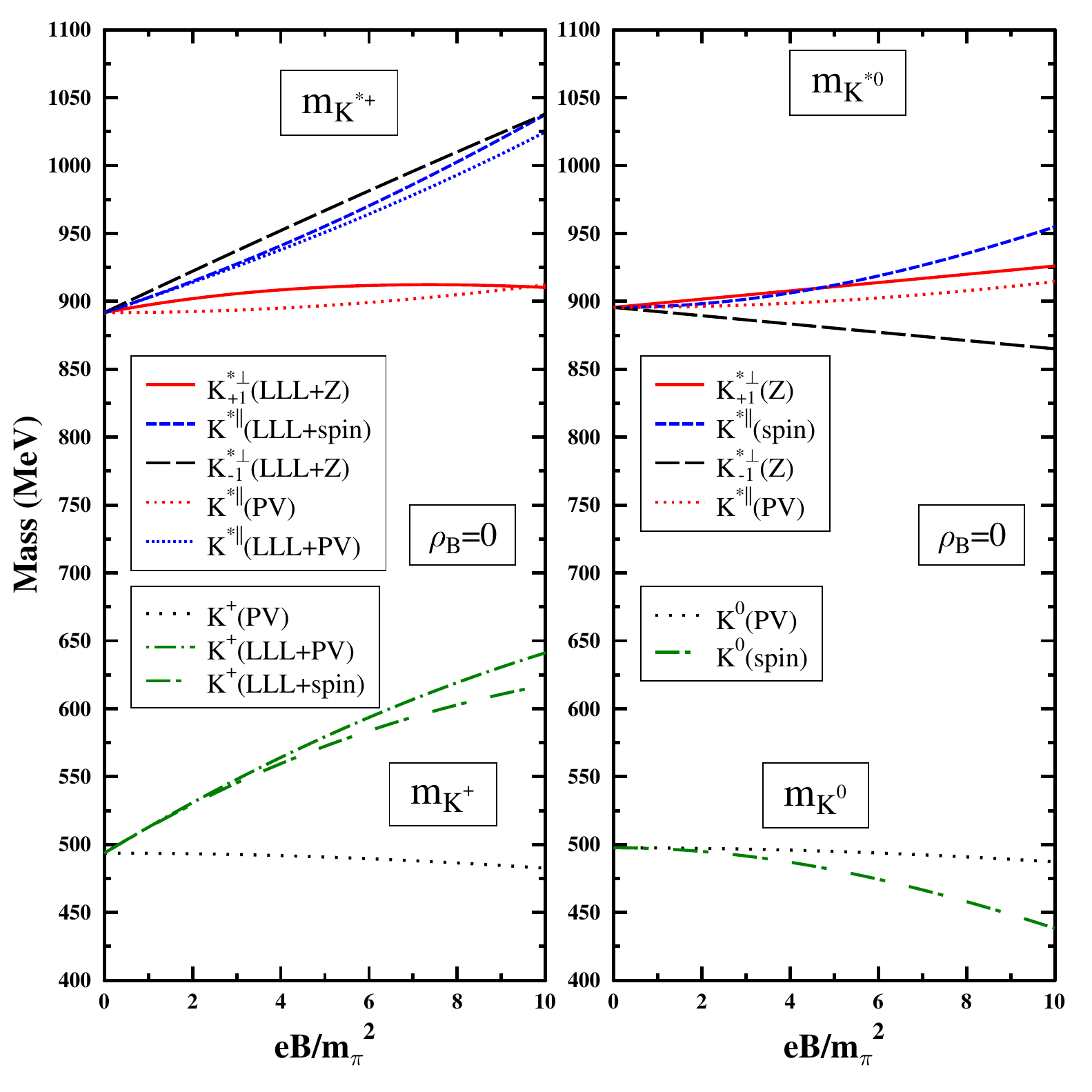}
\caption{Masses of vector ($K^*$) and pseudoscalar (K) meson plotted as a function of $eB/m_\pi^2$ at zero baryon density. Here `LLL' represents the lowest Landau level contribution, `PV' represents the PV mixing considered through effective Lagrangian, `spin' represents the PV mixing considered through spin-magnetic field interaction, and `Z' represents the anomalous Zeeman splitting. The three polarization states of vector $K^{*}$ meson \big($\big|1\,+1\big>, \big|1\,0\big>, \big|1\,-1\big>$\big) are written as $K^{*\perp}_{+1}, K^{*||}, K^{*\perp}_{-1}$.}
\label{mass_K*_mf_vacuum}
\end{figure}

\begin{figure}[hbt!]
\includegraphics[width=\linewidth]{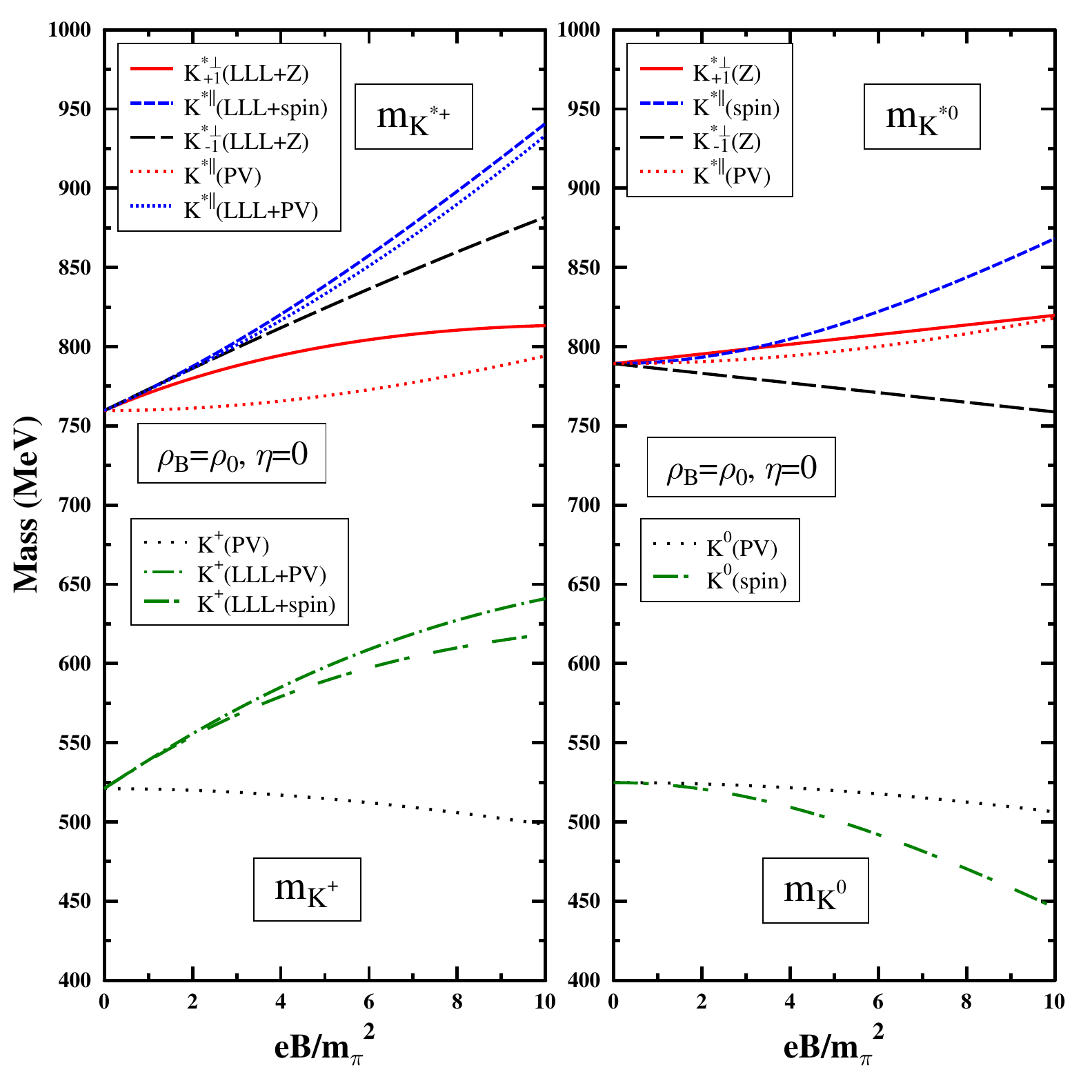}
\caption{Masses of vector ($K^*$) and pseudoscalar (K) meson plotted as a function of $eB/m_\pi^2$ at baryon density $\rho_B$= $\rho_0$ in isospin symmetric ($\eta=0$) nuclear matter. Here `LLL' represents the lowest Landau level contribution, `PV' represents the PV mixing considered through effective Lagrangian, `spin' represents the PV mixing considered through spin-magnetic field interaction, and `Z' represents the anomalous Zeeman splitting. The three polarization states of vector $K^{*}$ meson \big($\big|1\,+1\big>, \big|1\,0\big>, \big|1\,-1\big>$\big) are written as $K^{*\perp}_{+1}, K^{*||}, K^{*\perp}_{-1}$.}
\label{mass_K*_mf_rho0_eta0}
\end{figure}

\begin{figure}[hbt!]
\includegraphics[width=\linewidth]{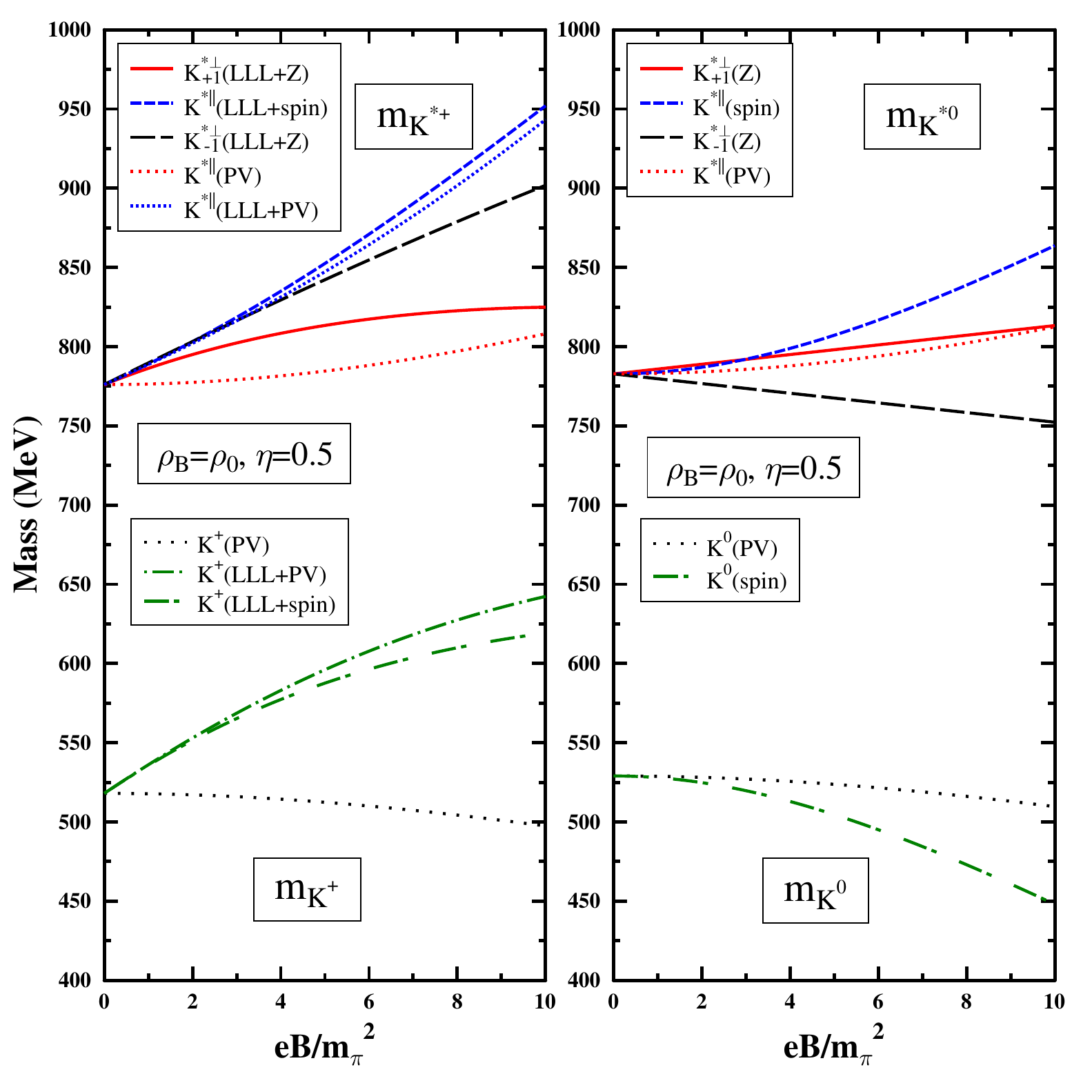}
\caption{Masses of vector ($K^*$) and pseudoscalar (K) meson plotted as a function of $eB/m_\pi^2$ at baryon density $\rho_B$= $\rho_0$ in isospin asymmetric ($\eta=0.5$) nuclear matter. Here `LLL' represents the lowest Landau level contribution, `PV' represents the PV mixing considered through effective Lagrangian, `spin' represents the PV mixing considered through spin-magnetic field interaction, and `Z' represents the anomalous Zeeman splitting. The three polarization states of vector $K^{*}$ meson \big($\big|1\,+1\big>, \big|1\,0\big>, \big|1\,-1\big>$\big) are written as $K^{*\perp}_{+1}, K^{*||}, K^{*\perp}_{-1}$.}
\label{mass_K*_mf_rho0_eta05}
\end{figure}

\subsection{Effects of Magnetic Field on $K^{*}$ Meson Masses}
We now discuss the effects of the magnetic field on the vector ($K^*$) and pseudoscalar ($K$) meson masses. First, due to lowest Landau level (LLL) contribution, the masses of the electrically charged mesons will be modified as given by equations (\ref{Landau_pseudo_2}) and (\ref{Landau_vector_2}). The charge neutral mesons will not undergo Landau quantization. We study the effects of the magnetic field at zero medium density as well as at medium density $\rho_B=\rho_0$. Furthermore, due to PV mixing effects incorporated through effective interaction term, the masses of the longitudinal part of vector meson $K^{*||}$ and pseudoscalar meson, both charged and neutral, are further modified as given by equation (\ref{M_PV_mixing}). The effects of PV mixing considered through the effective interaction term are labelled as `PV' in figures (\ref{mass_K*_mf_vacuum}), (\ref{mass_K*_mf_rho0_eta0}), and (\ref{mass_K*_mf_rho0_eta05}) and we conclude that the PV mixing effects alone are small. Due to PV mixing, there is a level repulsion between vector ($K^{*||}$) and pseudoscalar $(K)$ meson, and the mass of vector increases, while the mass of pseudoscalar $(K)$ decreases as a function of magnetic field. The effect of PV mixing increases even more at higher magnetic field values.

The effects of PV mixing incorporated through spin-magnetic field interaction, labelled as `spin', are also shown in figures (\ref{mass_K*_mf_vacuum}), (\ref{mass_K*_mf_rho0_eta0}), and (\ref{mass_K*_mf_rho0_eta05}) for medium density $\rho_B=0$, $\rho_0(\eta=0)$, and $\rho_0(\eta=0.5)$, respectively. The values are plotted after including the combined effects of lowest Landau level contribution and spin-magnetic field interaction term (shown as label `LLL + spin') for charged pseudoscalar $K$ meson and longitudinal polarization $K^{*||}$ of vector meson. The masses of transverse polarized states are also modified, due to spin-magnetic field interaction, in the presence of magnetic field due to unequal quark masses and/or charges for the $K^*$ meson. This is known as anomalous Zeeman splitting, shown as `Z', which is due to the non-vanishing intrinsic magnetic moment of the meson. This splitting contributes to both charged as well as neutral $K^*$ meson masses. The constituent quark masses are taken as $m_{u,d}= 330$ MeV and $m_s = 480$ MeV. The mass of vector $K^{*||}$ meson increases and that of pseudoscalar $(K)$ meson decreases when we consider only the spin mixing effects. For the charged vector $\big|1+1\big>$ state, there is a decrease in mass due to the spin interaction term, while the $\big|1 -1\big>$ state has a positive mass shift. However, for the neutral vector $\big|1 +1\big>$ \Big($\big|1 -1\big>$\Big) state, there is observed to be an increase (decrease) in the masses due to the spin interaction term which is due to the polarity of constituent quark/antiquark charges. The effects of medium density and isospin asymmetry on the pseudoscalar $K$ meson mass are calculated within a chiral $SU(3)$ model \cite{AM_EPJA45_169_2010, AM_EPJA55_107_2019}. The effects of magnetic field are observed to be quite appreciable and should be visible in the mass spectra of these particles and other observables in heavy ion collision experiments. The effects of PV mixing, considered through the spin-magnetic field interaction (-$\mu_i.B$) term, have also been studied on the masses of open bottom mesons and upsilon states in magnetized nuclear matter which, in turn, affects the decay width of $\Upsilon(4S)$ to $\bar{B}B$ pair appreciably \cite{AM_IJMPE31_2250060_2022}.

\subsection{Masses and Decay Widths of $K^*$ meson within a phenomenological model}
The in-medium mass of $K^*$ meson is calculated from the medium modified self-energy at one loop level. The effects of the medium are incorporated through in-medium $K$ meson masses, which are calculated in chiral $SU(3)$ model \cite{AM_EPJA45_169_2010, AM_EPJA55_107_2019}. The coupling parameter $g_{K^*}$ is fixed for various decay channels, from the vacuum decay width given by equation (\ref{DW_self_energy}). Then we fix the bare mass of $K^*$ meson using equation (\ref{mass_bare_mass_relation}) for a particular value of cut-off parameter $\Lambda_c$ by taking vacuum mass values for mesons. After fixing the bare mass  $\mathring{m}_{K^*}$ and coupling parameter $g_{K^*}$, we can calculate the in-medium masses for $K^*$ meson from equation (\ref{mass_bare_mass_relation}) by inserting in-medium $K$ meson masses. The mass shifts for $K^{*+}$ (from $K^+ \pi^0$ and $K^0 \pi^+$ loop) and $K^{*0}$ (from $K^0 \pi^0$ and $K^+ \pi^-$ loop) mesons in the nuclear medium at zero magnetic fields, for various values of cut-off parameter, are given in table III(a) and III(b) respectively. We observe a positive mass shift of $27.8$ MeV for the charged $K^{*+}$ meson for isospin symmetric nuclear matter at density $\rho_B$ = $\rho_0$ for cut-off parameter $\Lambda_c$ = $1000$ MeV. A positive self energy indicates that the nuclear mean field provides repulsion to $K^{*+}$ meson. This mass shift can be compared with a mass shift of around $+40$ MeV in self-consistent scattering amplitude calculations and a mass modification of $+50$ MeV, at $\rho_0$, in the low density T$\rho$ approximation \cite{Cabrera_JPCS503_012017_2014}.
\begin{center}
\textbf{Table III(a): Vector $K^{*+}$ Meson Mass shifts (MeV) from $K\pi$ loop for  $eB$ = 0}
\vspace{0.3cm}
\begin{tabular}{ |c||c|c|c||c|c|c||c|c|c||c|c|c|}
 \hline
&\multicolumn{6}{|c||}{$\rho_B = \rho_0$}&\multicolumn{6}{|c|}{$\rho_B = 4\rho_0$}\\
\hline
&\multicolumn{3}{|c||}{$\eta=0$}&\multicolumn{3}{|c||}{$\eta=0.5$}&\multicolumn{3}{|c||}{$\eta=0$}&\multicolumn{3}{|c|}{$\eta=0.5$}\\
\hline
 $\Lambda_c$& $K^+ \pi^0$ & $K^0 \pi^+$ & $(K \pi)^+$  & $K^+ \pi^0$ & $K^0 \pi^+$ & $(K \pi)^+$ & $K^+ \pi^0$ & $K^0 \pi^+$ & $(K \pi)^+$ & $K^+ \pi^0$ & $K^0 \pi^+$ & $(K \pi)^+$ \\  
 \hline
 1000 & 18.0  & 9.8  & 27.8  & 11.9 & 11.8 & 23.7 & 5.3 & 6.0 & 11.3 & 3.6 & 8.7 &12.3 \\
 \hline
 2000 & 10.3  & 4.2   & 14.5  & 5.9 & 6.1 & 12.0 & 4.3 & 12.0 & 16.3 & 1.6 & 7.6 & 9.2 \\
\hline
 3000 &5.5   &2.6  & 8.1  & 6.0 & 5.0& 11.0 & 9.2 & 2.2 &11.4 &3.9 &10.9 & 14.8  \\
\hline 
4000 & 7.1  & 8.1 & 15.2  & 7.1 & 8.0 & 15.1 & 11.8 & 8.9 & 20.7 & 6.3 & 11.0 & 17.3\\
\hline
\end{tabular}
\end{center}

\begin{center}
\textbf{Table III(b): Vector $K^{*0}$ Meson Mass shifts (MeV) from $K\pi$ loop for $eB$ = 0}\\
\vspace{0.3cm}
\begin{tabular}{ |c||c|c|c||c|c|c||c|c|c||c|c|c|  }
\hline
&\multicolumn{6}{|c||}{$\rho_B = \rho_0$}&\multicolumn{6}{|c|}{$\rho_B = 4\rho_0$}\\
\hline
&\multicolumn{3}{|c||}{$\eta=0$}&\multicolumn{3}{|c||}{$\eta=0.5$}&\multicolumn{3}{|c||}{$\eta=0$}&\multicolumn{3}{|c|}{$\eta=0.5$}\\
\hline
 $\Lambda_c$ & $K^0 \pi^0$ & $K^+ \pi^-$ & $(K \pi)^0$ & $K^0 \pi^0$ & $K^+ \pi^-$ & $(K \pi)^0$ &$ K^0 \pi^0$ & $K^+ \pi^-$ & $(K \pi)^0$ & $K^0 \pi^0$ & $K^+ \pi^-$ & $(K \pi)^0$\\ 
 \hline
 1000 &1.5   &  15.6 &  17.1 & 3.4 &16.9 &20.3 &1.8 &19.5 &21.3 & 0.7&7.7 &8.4 \\
 \hline
 2000 & 0.8  & 7.7  & 8.5  & 3.8 &0.6 &4.4 &2.1 & 3.1& 5.2& 0.6&3.7 &4.3 \\
\hline
 3000 & 2.2  & 0.4  &  2.6 & 1.8 & 1.7 & 3.5& 6.1 & 2.3&8.4 &3.6 & 5.4& 9.0 \\
\hline 
4000 & 6.5  & 2.0  &  8.5 & 4.6 & 12.7 & 17.3 & 4.0 & 5.4& 9.4 &5.6 & 4.5&10.1 \\
\hline
\end{tabular}
\end{center}

The partial decay widths of various decay channels of vector $K^{*+}$ and $K^{*0}$ mesons to pseudoscalar kaon and pion are also given in Tables III(c) and III(d), where the effects of medium density and isospin asymmetry are studied. These are calculated using equation (\ref{DW_self_energy}) by considering the in-medium masses of $K^*$ as calculated from the self-energy loop and $K$ meson masses calculated within the chiral model \cite{AM_EPJA55_107_2019}. As the increase in $K$ meson masses is more as compared to vector $K^*$ meson masses, the in-medium decay width is seen to decrease when compared to its vacuum value. We also observe that the decay widths have very small dependency on the cut-off parameter $\Lambda_c$ which is an encouraging result. The wide range of cut-off parameter value were taken to account for the rough estimate made by comparing the vertex form factor $u(q^2)$ of \ref{form_factors} with the vertex form factor calculated in the ${}^3P_0$ model.
\begin{center}
\textbf{Table III(c): Vector $K^{*+}$ Meson Decay Width from $K\pi$ loop for $eB$ = 0}\\
\vspace{0.3cm}
\begin{tabular}{ |c||c|c|c||c|c|c||c|c|c||c|c|c|  }
 \hline
&\multicolumn{6}{|c||}{$\rho_B$ = $\rho_0$}&\multicolumn{6}{|c|}{$\rho_B$ = $4\rho_0$}\\
\hline
&\multicolumn{3}{|c||}{$\eta=0$}&\multicolumn{3}{|c||}{$\eta=0.5$}&\multicolumn{3}{|c||}{$\eta=0$}&\multicolumn{3}{|c|}{$\eta=0.5$}\\
\hline
 $\Lambda_c$& $K^+ \pi^0$ & $K^0 \pi^+$ & $(K\pi)^+$ & $K^+ \pi^0$ & $K^0 \pi^+$ & $(K\pi)^+$ &  $K^+ \pi^0$ & $K^0 \pi^+$ & $(K\pi)^+$ & $K^+ \pi^0$ & $K^0 \pi^+$ & $(K\pi)^+$ \\
 \hline
 1000 & 15.6  & 29.6  &  45.2 & 15.4 & 29.1 & 44.5  & 12.2 & 24.4 &  36.6 & 13.6 & 16.7 & 30.3 \\
 \hline
 2000 & 15.0  & 28.7  &  43.7 & 14.9 & 28.1 & 43.0 & 12.1 & 25.4 & 37.5 & 13.5 & 16.5 & 30.0 \\
\hline
 3000 & 14.6   & 28.4  & 43.0  & 14.9 & 27.9 & 42.8 & 12.5 & 23.8 & 36.3  & 13.7 & 17.0 & 30.7 \\
\hline 
4000 & 14.7 & 29.3 & 44.0 & 15.0  & 28.4 & 43.4 & 12.7 & 24.9 & 37.6  & 13.8 & 17.0 & 30.8 \\
\hline
\end{tabular}
\end{center}
\vspace{0.3cm}

\begin{center}
\textbf{Table III(d): Vector $K^{*0}$ Meson Decay Width from $K\pi$ loop for $eB$ = 0}\\
\vspace{0.3cm}
\begin{tabular}{|c||c|c|c||c|c|c||c|c|c||c|c|c|}
 \hline
&\multicolumn{6}{|c||}{$\rho_B = \rho_0$}&\multicolumn{6}{|c|}{$\rho_B = 4\rho_0$}\\
\hline
&\multicolumn{3}{|c||}{$\eta=0$}&\multicolumn{3}{|c||}{$\eta=0.5$}&\multicolumn{3}{|c||}{$\eta=0$}&\multicolumn{3}{|c|}{$\eta=0.5$}\\
\hline
 $\Lambda_c$& $K^0 \pi^0$ &$ K^+ \pi^-$ & $(K\pi)^0$ & $K^0 \pi^0$ & $K^+ \pi^-$ & $(K\pi)^0$ & $K^0 \pi^0$ & $K^+ \pi^-$ & $(K\pi)^0$ &$ K^0 \pi^0$ & $K^+ \pi^-$ & $(K\pi)^0$ \\  
 \hline
 1000 & 13.3  & 28.5  & 41.8  & 13.1 & 29.2 & 42.3 & 11.3 & 24.7 & 35.9 & 7.5 & 25.8 & 33.3 \\
 \hline
 2000 & 13.3  & 27.3  &  40.6 & 13.1 & 26.7 & 39.8 & 11.3 & 22.2 & 33.5 & 7.5 & 25.2 & 32.6\\
\hline
 3000 & 13.4  & 26.1  &  39.5 & 12.9 & 26.9 & 39.8  & 11.6 & 22.1 & 33.7 & 7.7 & 25.4 & 33.1  \\
\hline 
4000 & 13.7  & 26.4 & 40.1   & 13.2 & 28.6 & 41.7 & 11.4 & 22.6 &  34.0 & 7.8 & 25.3 & 33.1 \\
\hline
\end{tabular}
\end{center}

We also compute the effects of strong magnetic field on vector $K^*$ masses, through the self energy loop, from the pseudoscalar $K$ meson masses calculated after including the effects of Landau quantization ( for charged mesons only) and spin mixing (through $-\mu_i.B$ term) in the presence of magnetic field. These are tabulated in III(e) and III(f) at various values of cut-off parameter $\Lambda_c$. We observe that the modifications in $K^*$ meson masses due to self energy loop are small when compared with the direct effects of magnetic field like Landau quantization and spin-magnetic field interaction. 

\begin{center}
\textbf{Table III(e): Vector $K^{*+}$ Meson Masses from $K\pi$ loop at $\rho_B = 0$}\\
\vspace{0.3cm}
\begin{tabular}{ |c||c|c|c|c||c|c|c|c|  }
\hline
&\multicolumn{4}{|c||}{$K^+ \pi^0$}&\multicolumn{4}{|c|}{$K^0 \pi^+$}\\
\hline
\backslashbox{$eB/m_\pi^2$}{$\Lambda_c$}&  1000& 2000 &3000 & 4000 & 1000 & 2000 & 3000 & 4000 \\
 \hline
 0 &  891.67& 891.67& 891.67& 891.67& 891.67& 891.67& 891.67& 891.67\\
 \hline
 2 &  887.81 &  894.24 & 894.72  & 892.58  & 889.75  &  904.00 &  894.77 &922.53\\
\hline
 4 &  893.73 &  889.92 &  894.03 & 899.75  & 900.90  & 886.30  &894.24   &893.08\\
\hline 
6 & 891.73  & 891.70  & 889.53  &890.44   & 897.61  & 894.75  &  895.77 &893.40\\
\hline
8 & 894.41  & 894.14  & 899.05  &895.98   & 890.769  & 892.27  &891.2   &893.14\\
\hline
10 & 895.75  & 895.60  & 898.77  &  895.94 & 895.51  & 898.78  &888.47   &880.38\\
\hline

\end{tabular}
\end{center}

\vspace{0.3cm}

\begin{center}
\textbf{Table III(f): Vector $K^{*0}$ Meson Masses from $K\pi$ loop at $\rho_B = 0$}\\
\vspace{0.3cm}
\begin{tabular}{ |c||c|c|c|c||c|c|c|c|  }
\hline
&\multicolumn{4}{|c||}{$K^0 \pi^0$}&\multicolumn{4}{|c|}{$K^+ \pi^-$}\\
\hline
\backslashbox{$eB/m_\pi^2$}{$\Lambda_c$}&  1000& 2000 &3000 & 4000 & 1000 & 2000 & 3000 & 4000 \\ 
 \hline
 0 & 895.55& 895.55& 895.55& 895.55& 895.55& 895.55& 895.55& 895.55\\
 \hline
 2 & 883.93  &890.73   & 896.40  &899.95   &  897.67 &902.91   &898.48   &901.17 \\
\hline
 4 & 893.47  &896.23   &896.75   & 905.33  & 903.26  & 905.13  & 894.39  &895.69\\
\hline 
6 &  890.78 &891.09   & 901.23  & 903.09  &  899.86 & 894.35  & 897.74  &896.41\\
\hline
8 &  890.47 & 888.51  &897.80   &898.71   &  904.42 & 901.33  & 902.63  &904.85\\
\hline
10 &  893.10 &891.30   &893.76   &899.33   &  905.43 & 904.15  & 905.23  &903.89\\
\hline

\end{tabular}
\end{center}
  
\begin{figure}[htp]
\centering
\includegraphics[width=\linewidth]{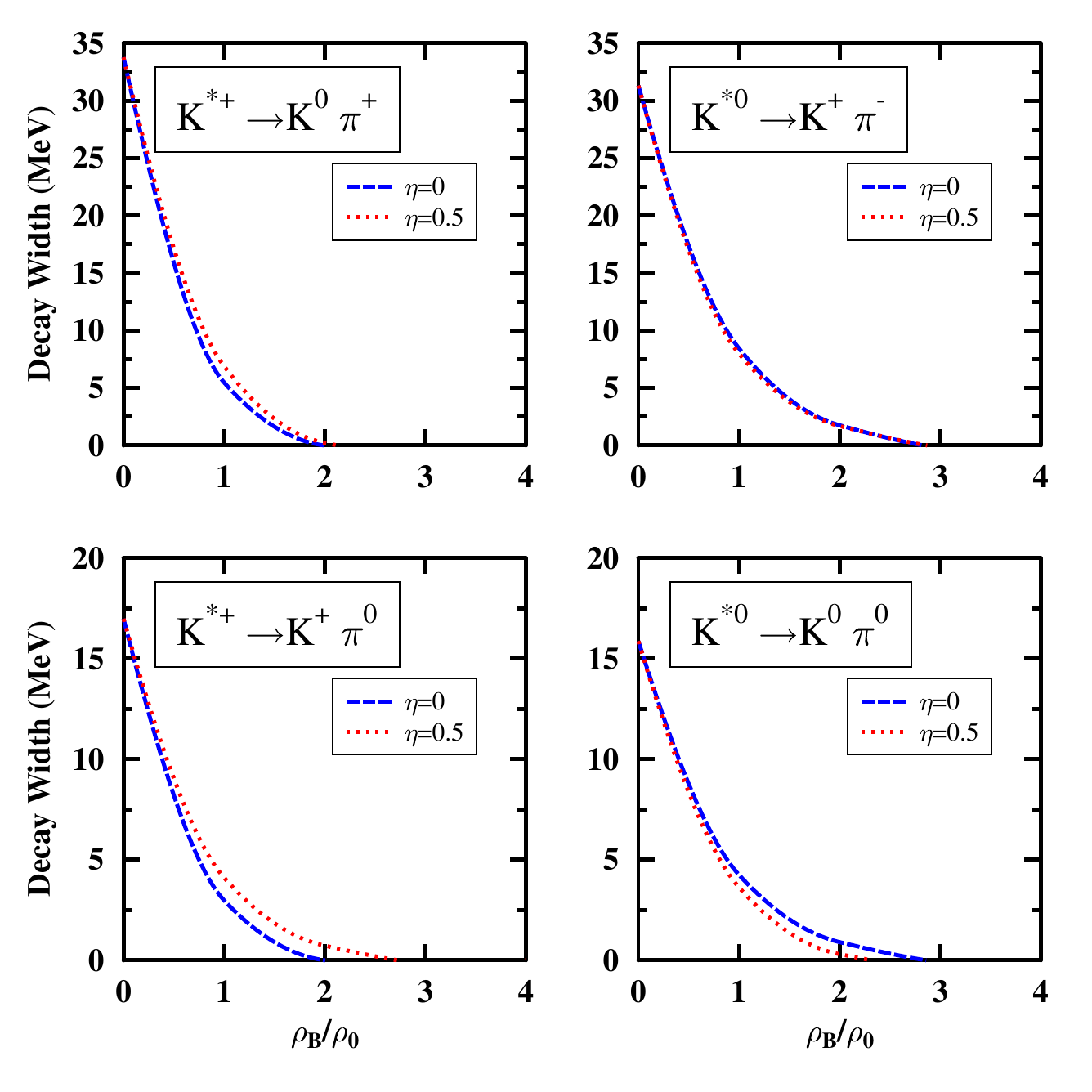}
\caption{Decay width of vector $K^*$ meson plotted as a function of relative baryon density ($\rho_B/\rho_0$) for various values of isospin asymmetry parameter $\eta$ for zero magnetic field case. }
\label{K*_DW_QCDSR}
\end{figure}

\subsection{In-medium Decay Width of Vector $K^*$ Meson within the ${}^3 P_0$ model}
We study the in-medium decay width of vector $K^*$ meson to two pseudoscalar mesons (kaon and pion), using the ${}^3 P_0$ model, from the mass modifications of vector $K^*$ meson and pseudoscalar kaon ($K$). The mass modifications for the pseudoscalar kaon have been studied using the chiral $SU(3)$ model \cite{AM_EPJA55_107_2019}. The vacuum masses for pseudoscalar mesons are taken to be $m_{K^+}=493.677$ MeV, $m_{K^0}=497.611$ MeV, $m_{\pi^{\pm}}= 139.57039$ MeV, $m_{\pi^{0}}=134.9768 $ MeV \cite{Zyla_PDG_2020}. By putting the vacuum values of decay widths $\Gamma (K^* \rightarrow K \pi)$ and vacuum masses for various decay channels, we find the coupling strength $\gamma$ related to the strength of ${}^3 P_0$ vertex of each channel individually. 

The coupling strength parameter $\gamma$ for the decays  $(K^{*+} \rightarrow K^+ \pi^0)$, $(K^{*+} \rightarrow K^0 \pi^+)$, $(K^{*0} \rightarrow K^0 \pi^0)$, $(K^{*0} \rightarrow K^+ \pi^-)$ comes out to be $0.1242963$, $0.1781849$, $0.1198312$, $0.1673111$ MeV, respectively for vacuum decay widths to be 16.98, 33.77, 15.87, and 31.31 MeV, respectively. We assume spherical harmonic oscillator potential for the wave functions of vector $K^*$ as well as for pseudoscalar $K$ and $\pi$ mesons. The harmonic oscillator strength parameter for pion, fitted from its charge radius squared value (0.4 ${\rm fm}^2$), is ($211$ MeV)$^{-1}$ \cite{SPM_PRD18_1673_1978,AM_IJMPE24_1550053_2015}. Then we find the harmonic oscillator strength parameter for the $K$ meson by assuming the ratio $\beta_K/\beta_{\pi}$ to be the same as the ratio of their charge radii, $(r_{\rm ch})_\pi/(r_{\rm ch})_K$. The charge radius for $K$ and $K^*$ mesons are taken to be $\big((r_{\rm ch})_K= 0.56\, {\rm fm}\big)$ and $\big((r_{\rm ch})_{K^*}= 0.74 \,{\rm fm}\big)$, which gives $\beta_K$ = ($238.3$ MeV)$^{-1}$ and $\beta_{K^*}$= ($184.84$ MeV)$^{-1}$ \cite{AM_EPJA57_98_2021}. By taking into account the mass modifications of vector $K^*$ meson calculated within the QCD sum rule approach and of pseudoscalar $K$ meson using the chiral $SU(3)$ model, we calculate the in-medium decay width for the decay $(K^* \rightarrow K \pi)$ for various sub-channels by using the ${}^3 P_0$ model. 

The decay width for the $K^*$ meson for various decay channels is plotted in figure (\ref{K*_DW_QCDSR}) as a function of relative medium density ($\rho_B / \rho_0$) for different isospin asymmetry parameter ($\eta=0,0.5$) and is also given in table IV. We observed that the decay width for each decay channel decreases as the medium density is increased. This is because the mass for $K^*$ meson decreases as a function of density, while the mass of pseudoscalar kaon (both $K^+$ and $K^0$) increases with density as calculated within the chiral $SU(3)$ model. The effect of isospin asymmetry on the decay width is also more pronounced at higher density. The drop in decay width for the channels $(K^{*+} \rightarrow K^+ \pi^0)$ and $ (K^{*+} \rightarrow K^0 \pi^+)$ is observed to be more for isospin symmetric case ($\eta=0$) as compared to isospin asymmetric case ($\eta\neq0$), while for the $(K^{*0} \rightarrow K^+ \pi^-)$ and $ (K^{*0} \rightarrow K^0 \pi^0)$ decay channels, the drop is observed to be more for isospin asymmetric case ($\eta\neq0$). This is due to the fact that as the medium density is increased, the mass of charged $K^+$ meson is modified less for isospin asymmetric matter as compared to isospin symmetric matter, while neutral $K^0$ meson mass is modified greatly for isospin asymmetric matter \cite{AM_EPJA55_107_2019}.

\begin{center}
\textbf{Table IV(a): Decay Width of Vector $K^{*+}$ Meson}\\
\vspace{0.3cm}
\begin{tabular}{|p{2cm}||p{2cm}|p{2cm}|p{2cm}|p{2cm}|}
 \hline
&\multicolumn{2}{|c|}{$\Gamma (K^{*+} \rightarrow K^+ \pi^0)$ (MeV)}&\multicolumn{2}{|c|}{$\Gamma (K^{*+} \rightarrow K^0 \pi^+)$ (MeV)} \\
 \hline
 \hline
 density & $\eta=0$ & $\eta =0.5$ & $\eta=0$ & $\eta =0.5$\\
 \hline
 \hline
 $\rho_0$   & 2.98  & 4.11 &  5.48  &6.90 \\
 \hline
 2$\rho_0$   & 0.10  & 0.73 &  0.05  &0.23\\
 
 \hline
\end{tabular}
\end{center}

\begin{center}
\textbf{Table IV(b): Decay Width of Vector $K^{*0}$ Meson}\\
\vspace{0.3cm}
\begin{tabular}{|p{2cm}||p{2cm}|p{2cm}|p{2cm}|p{2cm}|}
\hline
&\multicolumn{2}{|c|}{$\Gamma (K^{*0} \rightarrow K^0 \pi^0)$ (MeV)}&\multicolumn{2}{|c|}{$\Gamma (K^{*0} \rightarrow K^+ \pi^-) $(MeV)} \\
 \hline
 \hline
 density& $\eta=0$ & $\eta =0.5$ & $\eta=0$ & $\eta =0.5$\\
 \hline
 \hline
 $\rho_0$   &  4.25 &3.63 &   8.45 & 7.99\\
 \hline
 2$\rho_0$   & 0.91  & 0.30 &   1.76 & 1.69\\
 
 \hline
\end{tabular}
\end{center}

\begin{figure}[htp]
\centering
\includegraphics[width=16.5cm, height=15cm]{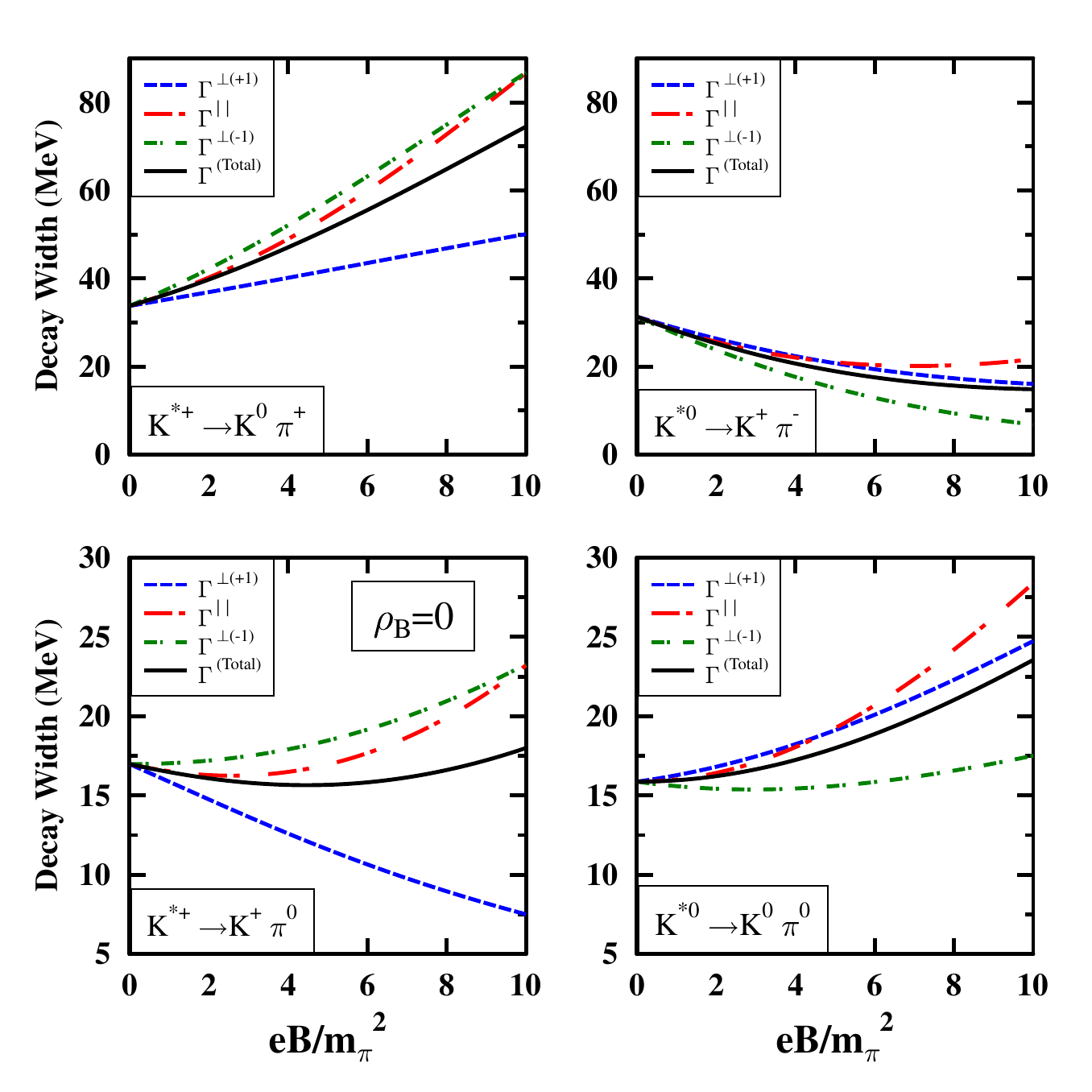}
\caption{Partial decay widths of vector $K^*$ meson plotted as a function of $eB/m_\pi^2$ at zero baryon density. The effects of Landau quantization and PV mixing, considered through spin-magnetic field interaction term, are included. The decay widths for different polarized states \big($\big|1\,+1\big>, \big|1\,0\big>, \big|1\,-1\big>$\big) are written as $\perp(+1)$, $|\,|$, and $\perp(-1)$, respectively.}
\label{K*_DW_mf_vacuum}
\end{figure}

\begin{figure}[htp]
\centering
\includegraphics[width=16.5cm, height=15cm]{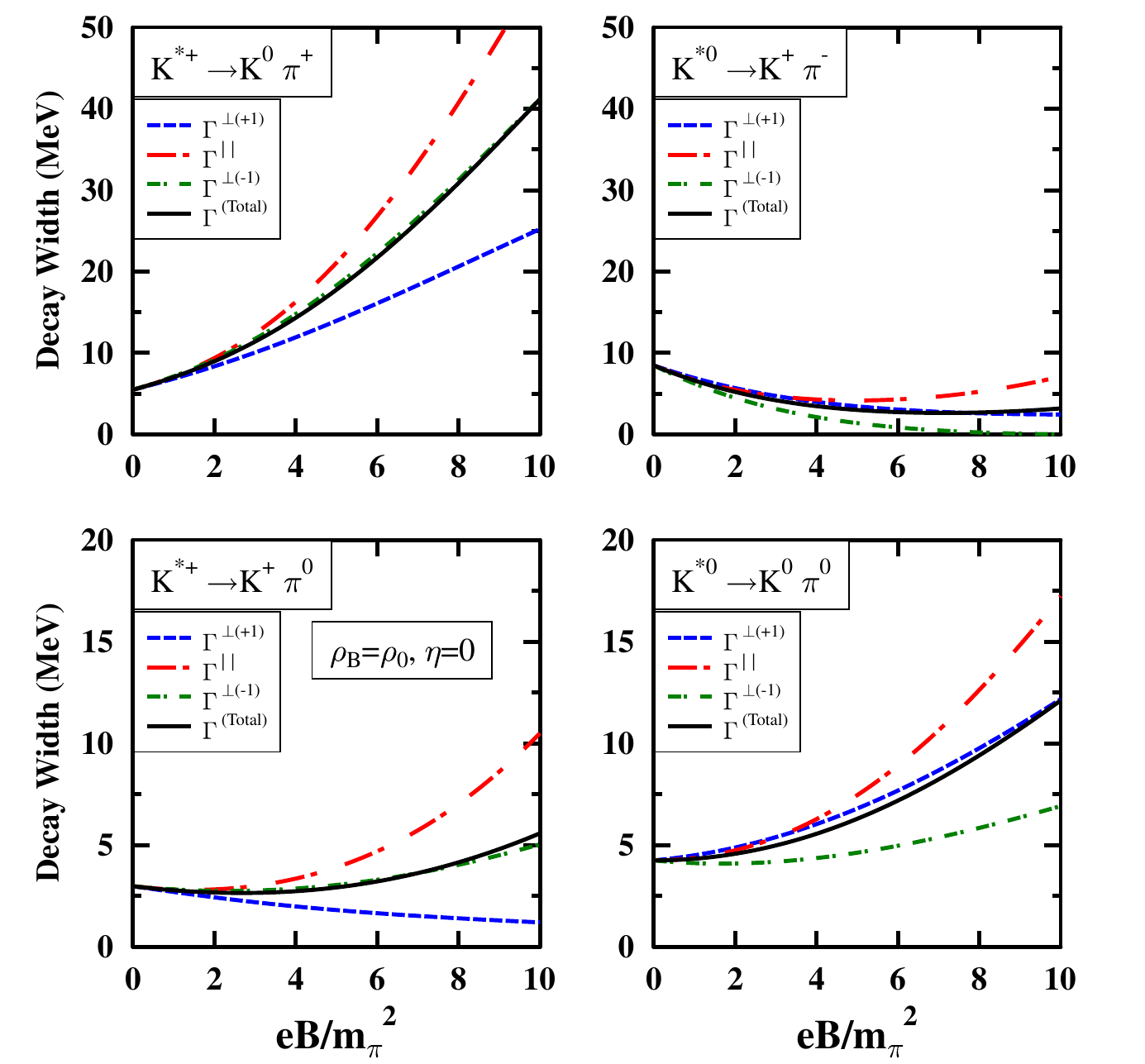}
\caption{Partial decay widths of vector $K^*$ meson plotted as a function of $eB/m_\pi^2$ at baryon density equal to nuclear matter saturation density ($\rho_0$) with $\eta=0$. The effects of Landau quantization and PV mixing, considered through spin-magnetic field interaction term, are included. The decay widths for different polarized states \big($\big|1\,+1\big>, \big|1\,0\big>, \big|1\,-1\big>$\big) are written as $\perp(+1)$, $|\,|$, and $\perp(-1)$, respectively.}
\label{K*_DW_mf_rho0_eta0}
\end{figure}

\begin{figure}[htp]
\centering
\includegraphics[width=16.5cm, height=15cm]{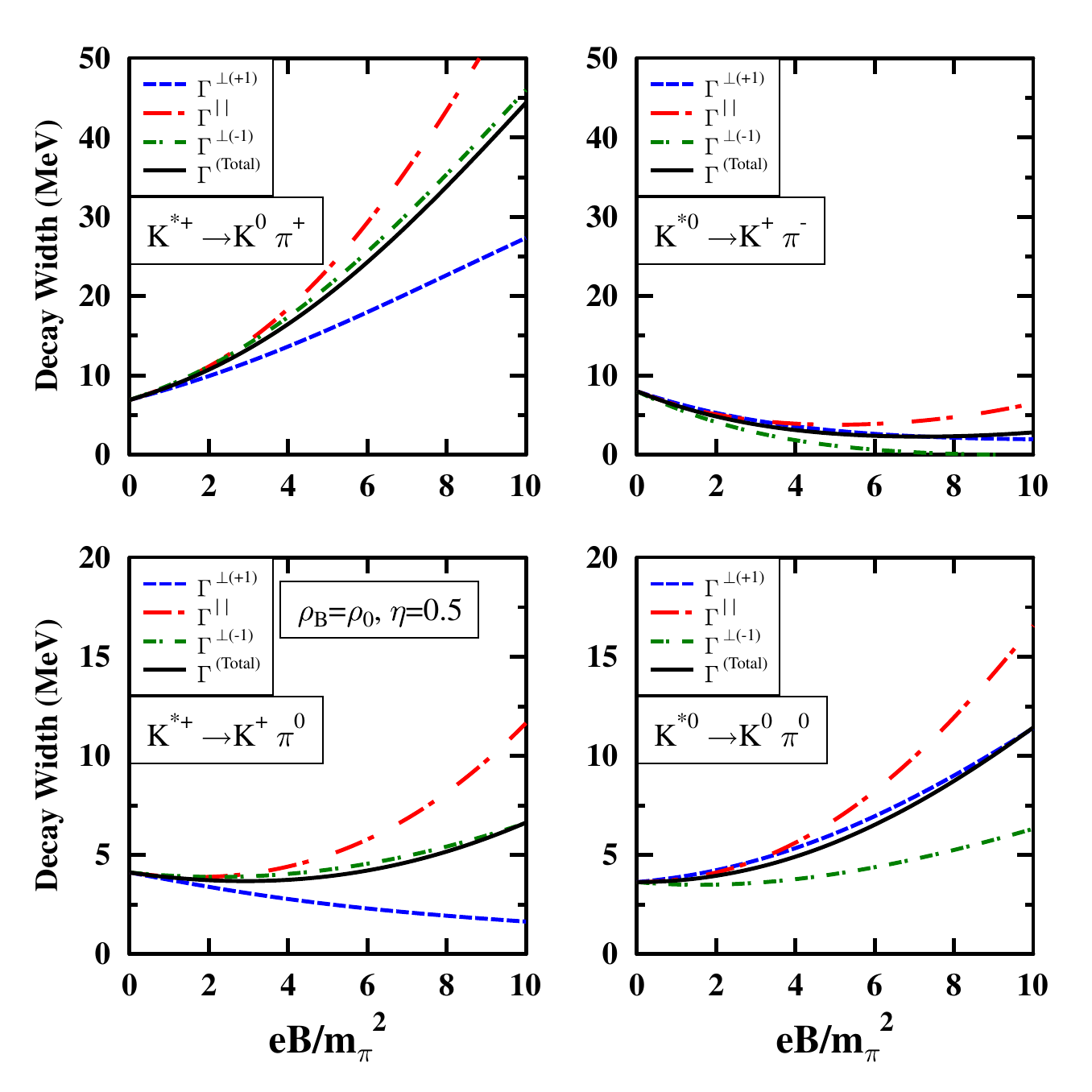}
\caption{Partial decay width of vector $K^*$ meson plotted as a function of $eB/m_\pi^2$ at baryon density equal to nuclear matter saturation density ($\rho_0$) with $\eta=0.5$. The effects of Landau quantization and PV mixing, considered through spin-magnetic field interaction term, are included. The decay widths for different polarized states \big($\big|1\,+1\big>, \big|1\,0\big>, \big|1\,-1\big>$\big) are written as $\perp(+1)$, $|\,|$, and $\perp(-1)$, respectively.}
\label{K*_DW_mf_rho0_eta05}
\end{figure}

\begin{figure}[htp]
\centering
\includegraphics[width=16.5cm, height=15cm]{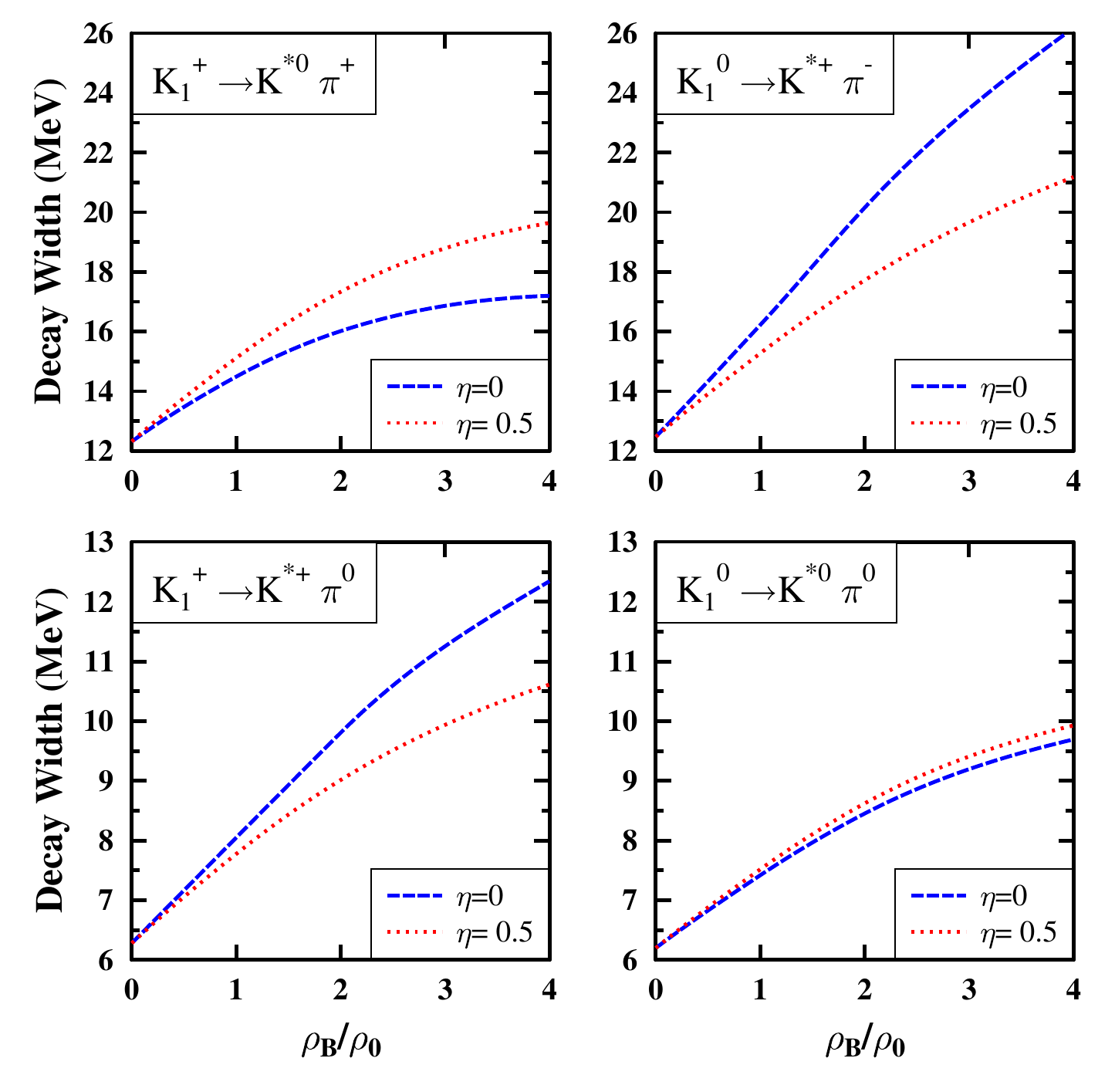}
\caption{Decay width of axial vector $K_1$ meson, plotted as a function of relative baryon density $(\rho_B/\rho_0)$, calculated within a phenomenological Lagrangian approach. }
\label{K1_DW_graph_phen}
\end{figure}

\begin{figure}[htp]
\centering
\includegraphics[width=16.5cm, height=15cm]{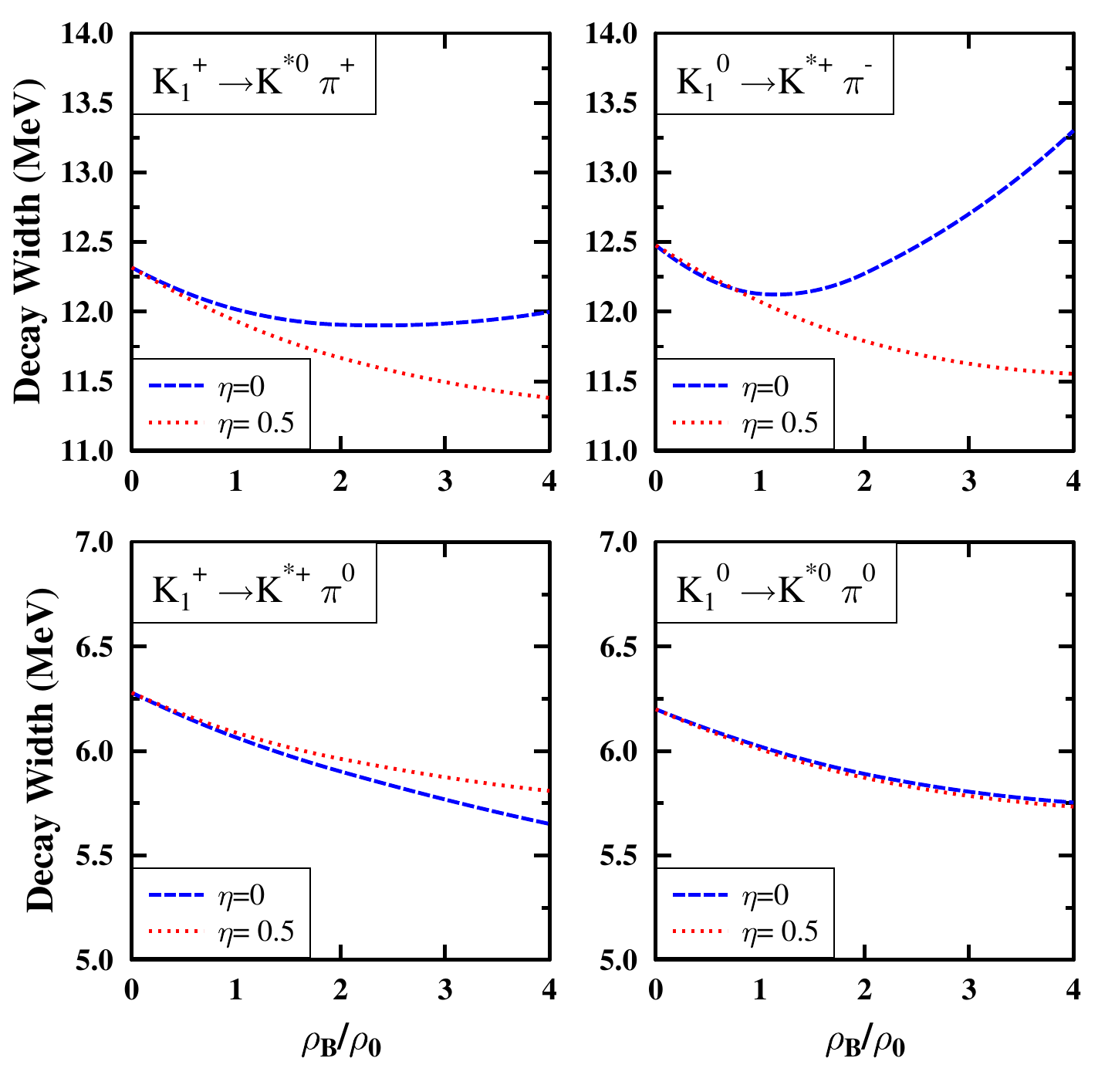}
\caption{Decay width of axial vector $K_1$ meson, plotted as a function of relative baryon density $(\rho_B/\rho_0)$, calculated within the ${}^3 P_0$ model.}
\label{K1_DW_graph_3P0}
\end{figure}

Furthermore, we have studied the effects of strong magnetic field on the decay widths of each decay channel by considering the effects from each polarization state individually. In Fig. (\ref{K*_DW_mf_vacuum}), the partial decay widths of vector $K^{*}$ meson are plotted as a function of $eB/m_\pi^2$ at zero baryon density after including the effects of Landau quantization and PV mixing (computed by using the spin mixing ($-\mu_i.B$) term). The qualitative behavior of decay width, as a function of magnetic field, reflects the medium dependence of the masses of mesons involved in that decay channel. As the neutral $K^{*0}$ vector meson does not undergo Landau quantization, the decay width modifications for its transverse polarized states arise due to medium modifications of vector meson masses due to Zeeman splitting and pseudoscalar meson mass due to PV mixing. On the other hand, the decay width modification for the longitudinal component of $K^{*0}$ arises due to medium modifications of masses of longitudinal part of vector meson and pseudoscalar meson due to PV mixing. In Figs. (\ref{K*_DW_mf_rho0_eta0}) and (\ref{K*_DW_mf_rho0_eta05}), the effects of magnetic field on the partial decay widths of $K^*$ are shown at nuclear medium density equal to nuclear matter saturation density $(\rho_0)$, in isospin symmetric $(\eta = 0)$ and asymmetric (with $\eta = 0.5)$ matter, respectively.

\subsection{In-medium Decay Width of Axial Vector $K_1$ Meson}
The two physical strange axial vector mesons $K_1(1270)$ and $K_1(1400)$ are the admixture of two strange members, $K_{1A}$ and $K_{1B}$, of two axial vector nonets. The larger branching ratio of $K_1(1400)$ to $K^* \pi$ as compared to $\rho K$ decay channel, and of $K_1(1270)$ to $\rho K$ as compared to $K^* \pi$ decay mode, indicates that there is a large mixing between $K_{1A}$ and $K_{1B}$ to give the physical mesons \cite{Roca_PRD70_094006_2004}. We have studied the in-medium $K_1 \rightarrow K^*\pi$ decay width from the medium modifications of the $K_1$ and $K^*$ meson masses, calculated within the QCD sum rule approach. In a phenomenological approach, the decay width is given by equation (\ref{DW_K1_phen2}). The parameters $\Tilde{F}$ and $\Tilde{D}$ are fitted in \cite{Roca_PRD70_094006_2004} from the various observed decay channels of $K_1(1270)$ and $K_1(1400)$ mesons as $1400$ MeV and $-1250$ MeV respectively. This fitting leads to the vacuum partial decay widths for the $K_1^{+}\rightarrow K^{*+}\pi^0$, $K_1^{+}\rightarrow K^{*0}\pi^+$, $K_1^{0}\rightarrow K^{*+}\pi^-$, and $K_1^{0}\rightarrow K^{*0}\pi^0$ decays as $6.2789, 12.3173, 12.4761$, and $6.1998$ MeV, respectively. The in-medium decay width is observed to increase as the medium density is increased as shown in figure (\ref{K1_DW_graph_phen}). This is due to a larger relative decrease in the $K^*$ meson mass as compared to the relative mass decrease for the $K_1$ meson within the QCDSR approach. As the density of the nuclear medium increases, we observed a significant increase of the effects of isospin asymmetric matter.

 Moreover, the $K_1$ meson decay width is also studied using the ${}^3 P_0$ model. The properties of strange axial vector mesons are still not very clear\cite{Guo_NPR36_125_2019} and we take the vacuum partial decay widths to be the same as calculated within the phenomenological approach. The decay widths of an axial vector ($A_1 \rightarrow \rho\pi$) are also studied in a phenomenological approach in reference \cite{Parui_arXiv_2209_02455}. The main idea here is to analyze the qualitative behavior of the strange axial-vector mesons. In figure (\ref{K1_DW_graph_3P0}), we have plotted the partial decay widths in nuclear medium calculated in ${}^3 P_0$ model. The variation trends in the decay width are caused by the interplay between the mass modifications of the involved $K^* (K^{*+}$ or $K^{*0})$ and $K_1(K_1^+$ or $K_1^0)$ mesons. There is observed to be a smaller variation as a function of medium density when compared with the results of the phenomenological approach.

\section{Summary}
\label{summary_section}
In the present work, we have studied the in-medium masses of vector $K^*$ meson and axial vector $K_1$ meson in isospin asymmetric nuclear matter, using the QCD sum rule approach. The self-energy loop of $K^*$ meson is evaluated at one loop level from the medium modifications of the kaon masses. In the presence of strong magnetic field, the effects of Landau quantization and PV mixing (through effective interaction vertex as well as spin-magnetic field interaction term) are also included for the $K^*$ meson and are found to be quite significant. The decay widths of various decay channels for the $K^* \rightarrow K\pi$ decay are also studied by using the ${}^3 P_0$ model. Also, the partial decay widths for the $K_1 \rightarrow K^*\pi$ decay are analyzed within a phenomenological approach as well as within the ${}^3 P_0$ model. In QCDSR, the mass modifications in the medium occur due to medium modifications of the light quark condensates and scalar gluon condensates, which are calculated using a chiral effective Lagrangian approach within a chiral effective model. The effects of density are found to be more pronounced than the effects of isospin asymmetry of the nuclear medium. The isospin asymmetry effects are observed to be more pronounced at higher medium density.  The effects of PV mixing are observed to be larger at larger magnetic fields. The effects of medium density and/or magnetic field on decay widths are a reflection of medium modifications of the masses of mesons involved. This present analysis of the in-medium properties of open strange particles might find relevance in heavy-ion collision experiments in the Relativistic Heavy Ion Collider (RHIC) low-energy scan programme and the High Acceptance DiElectron Spectrometer (HADES) Collaboration at GSI, Darmstadt.


\begin{thebibliography}{50}
\bibitem{Hosaka_PPNP96_88_2017} A. Hosaka, T. Hyodo, K. Sudoh, Y. Yamaguchi, and S. Yasui, Prog. Part. Nucl. Phys. \textbf{96}, 88 (2017).

\bibitem{AM_PRC91_035201_2015} A. Mishra, Phys. Rev. C \textbf{91}, 035201 (2015).

\bibitem{AM_PRC100_015207_2019} A. Mishra, A. Kumar, P. Parui, and S. De, Phys. Rev. C \textbf{100}, 015207 (2019).

\bibitem{Hartnack_PR510_119_2012} C. Hartnack, H. Oeschler, Y. Leifels, E. L. Bratkovskaya, and J. Aichelin, Phys. Rep. \textbf{510}, 119 (2012).

\bibitem{Tolos_PPNP112_103770_2020} L. Tolos and L. Fabbietti, Prog. Part. Nucl. Phys. \textbf{112}, 103770 (2020).

\bibitem{Kaplan_PLB175_57_1986} D. B. Kaplan and A. E. Nelson, Phys. Lett. B \textbf{175}, 57 (1986); A. E. Nelson and D. B. Kaplan, Phys. Lett. B \textbf{192}, 193 (1987).

\bibitem{AM_PRC70_044904_2004} A. Mishra, E. L. Bratkovskaya, J. Schaffner-Bielich, S. Schramm, and H. St\"ocker, Phys. Rev. C \textbf{70}, 044904 (2004).

\bibitem{AM_PRC78_024901_2008} A. Mishra, S. Schramm, and W. Greiner, Phys. Rev. C \textbf{78}, 024901 (2008).

\bibitem{AM_PRC74_064904_2006} A. Mishra and S. Schramm, Phys. Rev. C \textbf{74}, 064904 (2006).

\bibitem{AM_EPJA41_205_2009}  A. Mishra, A. Kumar, S. Sanyal, and S. Schramm, Eur. Phys. J. A \textbf{41}, 205 (2009).


\bibitem{Oset_NPA635_99_1998} E. Oset and A. Ramos, Nucl. Phys. A \textbf{635}, 99 (1998).

\bibitem{Ramos_NPA671_481_2000} A. Ramos and E. Oset, Nucl. Phys. A \textbf{671}, 481 (2000).

\bibitem{Koch_PLB337_7_1994} V. Koch, Phys. Lett. B \textbf{337}, 7 (1994).

\bibitem{Gubler_PLB767_336_2017} P. Gubler, T. Kunihiro, and S. H. Lee, Phys. Lett. B. \textbf{767}, 336 (2017).

\bibitem{Hatsuda_PRC46_R34_1992} T. Hatsuda and S. H. Lee, Phys. Rev. C \textbf{46}, R34 (1992).

\bibitem{AK_PRC82_045207_2010}  A. Kumar and A. Mishra, Phys. Rev. C \textbf{82}, 045207 (2010).

\bibitem{Parui_PRD106_114033_2022} P. Parui, S. De, A. Kumar, and A. Mishra, Phys. Rev. D \textbf{106}, 114033 (2022).

\bibitem{Ackleh_PRD54_6811_1996} E. S. Ackleh, T. Barnes, and E. S. Swanson, Phys. Rev. D \textbf{54}, 6811 (1996).

\bibitem{Micu_NPB10_521_1969} L. Micu, Nucl. Phys. B \textbf{10}, 521 (1969).

\bibitem{Yaouanc_PRD8_2223_1973} A. Le Yaouanc, L. Oliver, O. P\`ene, and J. C. Raynal, Phys. Rev. D \textbf{8}, 2223 (1973); ibid, Phys. Rev. D \textbf{9}, 1415 (1974); ibid, Phys. Rev. D \textbf{11}, 1272 (1975).


\bibitem{Barnes_PRD55_4157_1997} T. Barnes, F. E. Close, P. R. Page, and E. S. Swanson, Phys. Rev. D \textbf{55}, 4157 (1997).

\bibitem{Friman_PLB548_153_2002} B. Friman, S. H. Lee, and T. Song, Phys. Lett. B \textbf{548}, 153 (2002).

\bibitem{Yaouanc_PLB71_397_1977} A. Le Yaouanc, L. Oliver, O. P\`ene, and J. C. Raynal, Phys. Lett. B \textbf{71}, 397 (1977).

\bibitem{AM_IJMPE30_2150014_2021} A. Mishra and S. P. Misra, Int. J. Mod. Phys. E \textbf{30}, 2150014 (2021).

\bibitem{AM_EPJA57_98_2021} A. Mishra and S. P. Misra,  Eur. Phys. J. A \textbf{57}, 98 (2021).

\bibitem{Tolos_PRC82_045210_2010} L. Tolos, R. Molina, E. Oset, and A. Ramos, Phys. Rev. C \textbf{82}, 045210 (2010).

\bibitem{Ilner_PRC95_014903_2017} A. Ilner, D. Cabrera, C. Markert, and E. Bratkovskaya, Phys. Rev. C \textbf{95}, 014903 (2017).

\bibitem{Ilner_PRC99_024914_2019} A. Ilner, J. Blair, D. Cabrera, C. Markert, and E. Bratkovskaya, Phys. Rev. C \textbf{99}, 024914 (2019).

\bibitem{Kumar_EPJC79_403_2019} R. Kumar and A. Kumar, Eur. Phys. J. C \textbf{79}, 403 (2019).

\bibitem{Rapp_PNP65_209_2010} R. Rapp, D. Blaschke, and P. Crochet, Prog. Part. Nucl. Phys. \textbf{65}, 209 (2010).


\bibitem{Song_PLB792_160_2019} T. Song, T. Hatsuda, and S. H. Lee, Phys. Lett. B \textbf{792}, 160 (2019).

\bibitem{Tayduganov_PRD85_074011_2012} A. Tayduganov, E. Kou, and A. Le Yaouanc, Phys. Rev. D \textbf{85}, 074011 (2012).

\bibitem{Hatsuda_PRC52_3364_1995} T. Hatsuda, S. H. Lee, and H. Shiomi, Phys. Rev. C \textbf{52}, 3364 (1995).

\bibitem{Papazoglou_PRC59_411_1999} P. Papazoglou, D. Zschiesche, S. Schramm, J. Schaffner-Bielich, H. St\"ocker, and W. Greiner, Phys. Rev. C \textbf{59}, 411 (1999).

\bibitem{Zschiesche_PRC63_025211_2001} D. Zschiesche, P. Papazoglou, S. Schramm, J. Schaffner-Bielich, H. St\"ocker, and W. Greiner, Phys. Rev. C \textbf{63}, 025211 (2001).

\bibitem{Tuchin_AHEP2013_490495_2013} K. Tuchin, Adv. High Energy Phys. \textbf{2013}, 490495 (2013).

\bibitem{Skokov_IJMPA24_5925_2009} V. Skokov, A. Y. Illarionov, and V. Toneev, Int. J. Mod. Phys. A \textbf{24}, 5925 (2009).

\bibitem{Fukushima_PRD78_074033_2008} K. Fukushima, D. E. Kharzeev, and H. J. Warringa, Phys. Rev. D \textbf{78}, 074033 (2008).

\bibitem{Kharzeev_Springer_2013} D. Kharzeev, K. Landsteiner, A. Schmitt et al., \textit{Strongly Interacting Matter in Magnetic Fields}, Springer (2013).

\bibitem{Shovkovy_LNP871_13_2013}  I. A. Shovkovy, Lect. Notes Phys. \textbf{871}, 13 (2013).


\bibitem{Preis_LNP871_49_2013}  F. Preis, A. Rebhan, and A. Schmitt, Lect. Notes Phys. \textbf{871}, 49 (2013).

\bibitem{Gubler_PRD93_054026} P. Gubler, K. Hattori, S. H. Lee, M. Oka, S. Ozaki, and K. Suzuki, Phys. Rev. D \textbf{93}, 054026 (2016).

\bibitem{Cho_PRD91_045025_2015} S. Cho, K. Hattori, S. H. Lee, K. Morita, and S. Ozaki, Phys. Rev. D \textbf{91}, 045025  (2015); ibid, Phys. Rev. Lett. \textbf{113}, 172301 (2014).

\bibitem{AM_PRC102_045204_2020} A. Mishra and S. P. Misra, Phys. Rev. C \textbf{102}, 045204 (2020).

\bibitem{AM_IJMPE30_2150064_2021} A. Mishra and S. P. Misra, Int. J. Mod. Phys. E \textbf{30}, 2150064 (2021).

\bibitem{Alford_PRD88_105017_2013} J. Alford and M. Strickland, Phys. Rev. D \textbf{88}, 105017 (2013).

\bibitem{Yoshida_PRD94_074043_2016} T. Yoshida and K. Suzuki, Phys. Rev. D \textbf{94}, 074043 (2016).

\bibitem{Machado_PRD88_034009_2013} C. S. Machado, F. S. Navarra, E. G. de Oliveira, J. Noronha, and M. Strickland, Phys. Rev. D \textbf{88}, 034009 (2013).

\bibitem{AM_PRC69_024903_2004} A. Mishra, K. Balazs, D. Zschiesche, S. Schramm, H. St\"ocker, and W. Greiner, Phys. Rev. C \textbf{69}, 024903 (2004).

\bibitem{Weinberg_PR166_1568_1968} S. Weinberg, Phys. Rev. \textbf{166}, 1568 (1968).


\bibitem{Coleman_PR177_2239_1969}  S. Coleman, J. Wess, and B. Zumino, Phys. Rev. \textbf{177}, 2239 (1969); C. G. Callan, S. Coleman, J. Wess, and B. Zumino, Phys. Rev. \textbf{177}, 2247 (1969).

\bibitem{Bardeen_PR177_2389_1969} W. A. Bardeen and B. W. Lee, Phys. Rev. \textbf{177}, 2389 (1969).

\bibitem{Schechter_PRD21_3393_1980} J. Schechter, Phys. Rev. D \textbf{21}, 3393 (1980).

\bibitem{Lee_P72_97_2009} S. H. Lee and K. Morita, Pramana \textbf{72}, 97 (2009).

\bibitem{Morita_PRC77_064904_2008}  K. Morita and S. H. Lee, Phys. Rev. C \textbf{77}, 064904 (2008).

\bibitem{Klingl_NPA624_527_197} F. Klingl, N. Kaiser, and W. Weise, Nuclear Physics A \textbf{624}, 527 (1997).

\bibitem{Leupold_PRC64_015202_2001} S. Leupold, Phys. Rev. C \textbf{64}, 015202 (2001).

\bibitem{Shifman_NPB147_385_1979} M. A. Shifman, A. I. Vainshtein, and V. I. Zakharov, Nucl. Phys. B \textbf{147}, 385 (1979).

\bibitem{Reinders_PR127_1_1985} L. J. Reinders, H. Rubinstein, and S. Yazaki, Phys. Reports \textbf{127}, 1 (1985).

\bibitem{Kwon_PRC81_065203_2010} Y. Kwon, C. Sasaki, and W. Weise, Phys. Rev. C \textbf{81}, 065203 (2010).


\bibitem{Shifman_NPB147_448_1979} M. A. Shifman, A. I. Vainshtein, and V. I. Zakharov, Nucl. Phys. B \textbf{147}, 448 (1979).

\bibitem{Parui_arXiv_2209_02455} P. Parui and A. Mishra, arXiv:2209.02455v1 [hep-ph].

\bibitem{Chernodub_LNP871_143_2013} M. N. Chernodub, Lect. Notes Phys. \textbf{871}, 143 (2013); M. N. Chernodub, Phys. Rev. D \textbf{82}, 085011 (2010).

\bibitem{Iwasaki_EPJA57_222_2021} S. Iwasaki, M. Oka, and K. Suzuki, Eur. Phys. J. A \textbf{57}, 222 (2021).

\bibitem{Klingl_ZPA356_193_1996} F. Klingl, N. Kaiser, and W. Weise, Z. Phys. A \textbf{356}, 193 (1996).

\bibitem{Cobos-Martinez_PLB771_113_2017} J. J. Cobos-Martinez, K. Tsushima, G. Krein, and A. W. Thomas, Phys. Lett. B \textbf{771}, 113 (2017).

\bibitem{Itzykson_QFT_1980} C. Itzykson and J. B. Zuber, \textit{Quantum Field Theory}, Mc Graw-Hill, New York (1980).

\bibitem{Leinweber_PRD64_094502_2001} D. B. Leinweber, A. W. Thomas, K. Tsushima, and S. V. Wright, Phys. Rev. D \textbf{64}, 094502 (2001).

\bibitem{Krein_PLB697_136_2011} G. Krein, A. W. Thomas, and K. Tsushima, Phys. Lett. B \textbf{697}, 136 (2011).

\bibitem{Lee_PRC67_038202_2003} S. H. Lee and C. M. Ko, Phys. Rev. C \textbf{67}, 038202 (2003).


\bibitem{Roca_PRD70_094006_2004} L. Roca, J. E. Palomar, and E. Oset, Phys. Rev. D \textbf{70}, 094006 (2004).

\bibitem{Suzuki_PRD47_1252_1993} M. Suzuki, Phys. Rev. D \textbf{47}, 1252 (1993).

\bibitem{Zyla_PDG_2020} P. A. Zyla et al. (Particle Data Group), Prog. Theor. Exp. Phys. 2020, 083C01 (2020) and 2021 update.

\bibitem{AM_EPJA45_169_2010} A. Mishra, A. Kumar, S. Sanyal, V. Dexheimer, and S. Schramm, Eur. Phys. J. A \textbf{45}, 169 (2010).

\bibitem{AM_EPJA55_107_2019} A. Mishra, A. K. Singh, N. S. Rawat, and P. Aman, Eur. Phys. J. A \textbf{55}, 107 (2019).

\bibitem{AM_IJMPE31_2250060_2022} A. Mishra and S. P. Misra, Int. J. Mod. Phys. E \textbf{31}, 2250060 (2022).

\bibitem{Cabrera_JPCS503_012017_2014} D. Cabrera, L. M. Abreu, E. Bratkovskaya, A. Ilner, F. J. Llanes-Estrada, A. Ramos, L. Tolos, and J. M. Torres-Rincon, J. Phys. Conf. Ser. \textbf{503}, 012017 (2014).

\bibitem{SPM_PRD18_1673_1978} S. P. Misra, Phys. Rev. D \textbf{18}, 1673 (1978).

\bibitem{AM_IJMPE24_1550053_2015} A. Mishra, S. P. Misra, and W. Greiner, Int. J. Mod. Phys. E \textbf{24}, 1550053 (2015).

\bibitem{Guo_NPR36_125_2019} Peng-Fei Guo, D. Wang, and Fu-Sheng Yu, Nuclear Physics Review \textbf{36}, 125 (2019).






\end{thebibliography}
\end{document}